\documentclass[pdflatex,sn-mathphys-num]{sn-jnl}

\usepackage{graphicx}%
\usepackage{multirow}%
\usepackage{amsmath,amssymb,amsfonts}%
\usepackage{amsthm}%
\usepackage{mathrsfs}%
\usepackage[title]{appendix}%
\usepackage{xcolor}%
\usepackage{textcomp}%
\usepackage{manyfoot}%
\usepackage{booktabs}%
\usepackage{algorithm}%
\usepackage{algorithmicx}%
\usepackage{algpseudocode}%
\usepackage{listings}%
\theoremstyle{thmstyleone}

\theoremstyle{thmstyletwo}

\theoremstyle{thmstylethree}

\begin{document}

\title[ANTIOCH-VS]{ANTIOCH-VS: From Sequence to Hits with Modular, Open-Source Ensemble Virtual Screening}

\author*[1]{\fnm{Hale S.} \sur{Hasdemir}}\email{hhasdemir@som.umaryland.edu}

\author[2,3]{\fnm{Venkata Sai Sreyas} \sur{Adury}}\email{venkata@umd.edu}

\author*[4,3,1]{\fnm{Pratyush} \sur{Tiwary}}\email{ptiwary@umd.edu}

\affil*[1]{\orgname{University of Maryland Institute for Health Computing}, \orgaddress{\city{Bethesda}, \postcode{20852}, \state{Maryland}, \country{United States}}}

\affil[2]{\orgdiv{Chemical Physics Program}, \orgname{University of Maryland}, \orgaddress{\city{College Park}, \postcode{20742}, \state{Maryland}, \country{United States}}}

\affil[3]{\orgdiv{Institute for Physical Science and Technology}, \orgname{University of Maryland}, \orgaddress{\city{College Park}, \postcode{20742}, \state{Maryland}, \country{United States}}}

\affil[4]{\orgdiv{Department of Chemistry and Biochemistry}, \orgname{University of Maryland}, \orgaddress{\city{College Park}, \postcode{20742}, \state{Maryland}, \country{United States}}}

\abstract{Structure-based virtual screening needs a receptor conformation that fits the ligand class, but for most targets no ligand-bound crystal is available---and a single predicted model or an apo structure often misses the binding-competent state. We present ANTIOCH-VS, an open-source framework that runs an entire \textit{in silico} screening campaign, from a protein sequence to a ranked shortlist of candidate binders, as one plug-and-play workflow. It first builds a conformational ensemble by reduced-MSA subsampling with ColabFold, then uses AlphaFold2-RAVE (AF2RAVE)---short molecular-dynamics simulations with State Predictive Information Bottleneck analysis---to pick out docking-ready, Boltzmann-weighted metastable states. It then detects the binding pocket, prepares receptors and ligands, docks the library across the ensemble with DOCK6 or AutoDock Vina, builds a consensus ranking, and triages the top hits by physics-based metadynamics (CTMD). We benchmark it on two very different targets, JAK2 kinase and ADRB2 GPCR, using their DUD-E benchmark sets of known binders and property-matched decoys. Starting from sequence alone---no ligand-bound crystal, no manual intervention---the AF2RAVE ensembles gave top-20 shortlists strongly enriched for true binders: a 20\% hit rate for JAK2 and 60\% for ADRB2 (random $\approx$1--2\%), on par with ligand-bound crystals and, for the kinase, above single AF2 models and an apo crystal. The full pipeline runs end to end on a workstation or HPC from a single command, or each module can be run on its own with user-supplied inputs. ANTIOCH-VS is freely available at \url{https://github.com/tiwarylab/antioch-vs}.}

\keywords{Structure-based virtual screening, Protein conformational ensembles, Molecular docking, Metadynamics}

\maketitle

\section{Introduction}\label{sec1}
Structure-based virtual screening is widely used in early-stage drug discovery, but a practical campaign comprises multiple interdependent stages, including receptor and ligand preparation, binding-site and search-space definition, docking, post-docking evaluation, and prioritization for downstream simulation or experimental validation~\cite{glaab_building_2016}. Connecting these stages requires task-specific software, compatible molecular representations, reliable recording of intermediate inputs and outputs, and scalable execution across available computing resources. Although recent automated and containerized docking pipelines have improved accessibility and deployment, key tasks such as protein preparation and docking-grid definition can still require separate tools and manual intervention~\cite{minibaeva_easydock_2026}. Moreover, making informed choices at each stage demands domain expertise that contributes to an already significant barrier to entry. Consequently, end-to-end virtual-screening workflows remain challenging to independently devise, reproduce, transfer across computing environments, and extend to new receptor ensembles or docking engines.

The very first challenge one faces in such a campaign is the availability of a suitable structure for the target. Proteins exist as conformational ensembles, and individual members of an ensemble can differ substantially in the accessibility, geometry, and side-chain organization of ligand-binding pockets. This challenges conventional methods for structure-based drug discovery when a reliable ligand-bound crystal structure is unavailable. Recent advances in deep learning based protein structure prediction methods like AlphaFold2 (AF2)~\cite{jumper_highly_2021} and Boltz-2~\cite{passaro2025boltz2} rely on sequence alone to generate structure hypotheses. These methods typically predict just one most probable structure offering little structural diversity. An AF2 model selected by internal confidence metrics, or an available apo crystal structure, may capture a stable ligand-free state but miss the ligand-compatible conformation required for docking. A systematic redocking evaluation confirmed that AF2 confidence scores did not reliably predict docking performance, even when predicted backbones closely match experimental structures~\cite{holcomb_evaluation_2023}. For receptors with substantial flexibility, ensemble docking may be necessary, as no single structure---even a reliable ligand-bound crystal---can capture the binding modes of all ligands~\cite{gu_hierarchical_2025}.\\
It is possible to generate a more diverse ensemble of protein structures using methods such as MSA-subsampling~\cite{delalamo_sampling_2022}. However, this often results in many more hallucinations that make predictions inaccurate for protein classes far from their training distribution unless supported by physics-based approaches~\cite{aranganathan_boltzmann_2025}.

AlphaFold2-RAVE (AF2RAVE) addresses these limitations by combining diverse AF2 predictions with molecular-dynamics (MD) simulations and State Predictive Information Bottleneck (SPIB) analysis to identify and Boltzmann-rank metastable conformational states~\cite{vani_alphafold2-rave_2023,teng_af2rave_2025}. Subsequent applications sampled metastable kinase conformations---including the classical DFG-out state relevant to type II inhibitor binding---accurately enough to dock type II inhibitors, and recovered how the relative stabilities of the DFG-in and DFG-out states shift with kinase sequence \cite{gu_empowering_2024,vani_exploring_2024}. Hierarchical AF2RAVE further extended this approach by separately resolving global backbone motions and local binding-pocket side-chain metastability. Applied to Ca$^{2+}$--S100B, it generated holo-like receptor conformations that improved ligand pose recovery relative to standard AlphaFold2 models and outperformed both standard AF2 and available X-ray receptor structures in virtual-screening enrichment tests~\cite{gu_hierarchical_2025}. Collectively, these studies establish AF2RAVE as a promising approach for generating and prioritizing docking-relevant receptor conformations when experimentally determined ligand-bound structures are unavailable.

Modern docking engines provide efficient and increasingly capable methods for virtual screening. AutoDock Vina combines an empirical scoring function with efficient optimization and multithreaded execution~\cite{trott_autodock_2010}; later releases added expanded scoring options, batch docking, and Python bindings to support programmatic workflows~\cite{eberhardt_autodock_2021}. GPU implementations have further accelerated Vina and its derivatives for large-scale screens while maintaining comparable or improved docking performance~\cite{ding_vina-gpu_2023,tang_vina-gpu_2024}. Likewise, recent DOCK6 developments introduced hierarchical traversal of precomputed ligand conformations to expand access to large chemical libraries, including a demonstrated screen of 23 million compounds~\cite{balius_dock_2024}. Despite these advances, these programs are principally docking engines: they do not, by themselves, define a complete and reproducible workflow spanning receptor and ligand preparation, search-space specification, batch execution, result extraction, enrichment evaluation, ensemble consensus ranking, and handoff of selected receptor--ligand complexes to molecular-dynamics simulations. Consequently, researchers often rely on project-specific scripts and manual file transformations to connect these stages, which can make workflows difficult to reproduce, transfer across computing environments, and adapt to new targets or docking engines.

\begin{figure}[htbp]
    \centering
    \includegraphics[width=\textwidth]{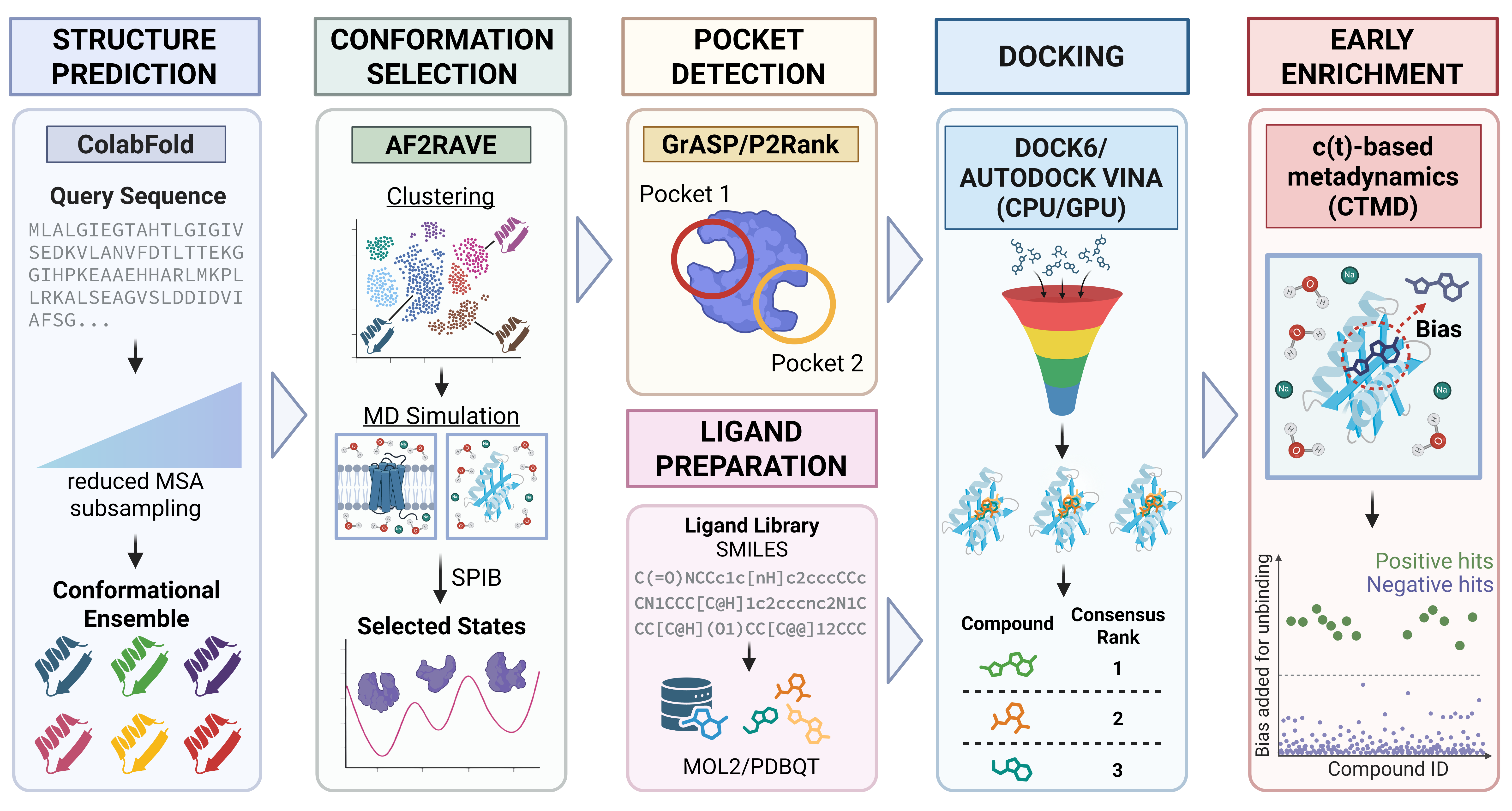}
    \caption{\textbf{The ANTIOCH-VS pipeline.} Starting from a target sequence, ANTIOCH-VS proceeds through five modules (left to right). \textbf{Structure prediction}: ColabFold with reduced-MSA subsampling generates a diverse conformational ensemble. \textbf{Conformation selection}: AF2RAVE clusters the ensemble, seeds unbiased MD, and applies State Predictive Information Bottleneck (SPIB) analysis to select a few Boltzmann-ranked metastable states for docking. \textbf{Pocket detection}: GrASP and P2Rank locate candidate binding pockets when none is known in advance; a companion ligand-preparation step converts library SMILES into docking-ready inputs (MOL2 for DOCK6, PDBQT for Autodock Vina). \textbf{Docking}: each ensemble receptor is screened against the ligand library with DOCK6 or AutoDock Vina (CPU/GPU), and the per-receptor rankings are combined into a single consensus ranking. \textbf{Early enrichment}: top-ranked complexes are triaged by $c(t)$-based metadynamics (CTMD), a physics-based post-docking step that separates stably bound (positive) from weekly bound (negative) hits before costlier free-energy calculations or experimental validation.}
    \label{fig:pipeline}
\end{figure}

In this work, we present ANTIOCH-VS (Automated, Networked Toolkit for In-silico Open-source Conformational-ensemble Hit-discovery --- Virtual Screening), a modular and reproducible framework for ensemble virtual screening and post-docking preparation.
ANTIOCH-VS streamlines virtual screening pipelines into simple modules that deal with all of the following in a single plug-and-play containerized framework (Fig.~\ref{fig:pipeline}):
\begin{enumerate}
    \item  Structure hypothesis generation using pretrained models like AF2 and even allows user supplied receptor structures
    \item Receptor and ligand preparation pipelines using a combination of PDBFixer~\cite{eastman_openmm_2017}, UCSF Chimera~\cite{pettersen_ucsf_2004}, and AmberTools~\cite{case_ambertools_2023} for receptor preparation, Meeko~\cite{santos-martins_meeko_2025} for receptor PDBQT generation, and a universal molecular descriptor (UMD) format for converting ligand libraries into docking-ready inputs (Mol2 for DOCK6, PDBQT for Vina)
    \item Docking support for multiple engines, with AutoDock Vina~\cite{trott_autodock_2010,eberhardt_autodock_2021} and DOCK 6~\cite{balius_dock_2024} supported out of the box
    \item Automatic consensus ranking across the receptor ensemble, with enrichment-based evaluation for retrospective benchmarks
    \item Staging of selected receptor-ligand complexes for c(t)-based metadynamics (CTMD), a physics-based post-docking hit-triaging approach that prioritizes stable binding candidates before more computationally intensive free-energy calculations or experimental validation~\cite{adury_early-enrichment_2026}.
\end{enumerate}

By combining configuration-driven execution with HPC- and container-compatible workflows, ANTIOCH-VS enables researchers to assemble, evaluate, and extend virtual-screening campaigns without relying on a proprietary end-to-end platform.
Starting directly from sequence, with no ligand-bound crystal structure and no manual intervention, the pipeline yields top-20 shortlists strongly enriched for true binders (random expectation $\approx$1--2\%). For Janus kinase 2 (JAK2), the AF2RAVE-selected ensemble of 5 protein conformations reaches a 20\% hit rate (4/20) both after docking and after CTMD triage---above an identical docking process using AF2 structures directly (5\% docking, 15\% after CTMD) and on par with the reference ligand-bound crystal (3LPB, 20\% docking, 35\% after CTMD). For $\beta_2$-adrenergic receptor (ADRB2), the sequence-based ensembles are already strongly enriched: AF2RAVE recovers 60\% (12/20) and AF2 structures show 100\% (20/20), matching or exceeding the ligand-bound crystals (best 95\%). ANTIOCH-VS generalizes to any protein target, and its modular design permits straightforward substitution of individual components, such as alternative MD engines or docking backends without restructuring the workflow. We believe that the simplicity and intervention-free nature of this pipeline will make it a free and open-source solution for end-to-end (sequence to prospective hit) virtual screening campaigns.

\section{Results}\label{sec2}

ANTIOCH-VS assembles the stages of a virtual-screening campaign---conformational-ensemble generation, receptor and ligand preparation, docking across the receptor ensemble, consensus ranking of the results, and physics-based post-docking triage---into one configuration-driven workflow that can be run end-to-end without intervention (Fig.~\ref{fig:pipeline}). To establish that it works end to end, we ran two retrospective campaigns on targets chosen to exercise different parts of the pipeline: the soluble kinase JAK2 and the membrane protein ADRB2, a G-protein-coupled receptor (GPCR). Both are standard benchmarks with a Directory of Useful Decoys, Enhanced (DUD-E)~\cite{mysinger_dude_2012} active/decoy set and several ligand-bound crystal structures for comparison. We describe each campaign at a conceptual level here, deferring system-specific parameters (feature selections, MSA depths, simulation and docking settings) to Methods and the SI. The ANTIOCH-VS pipeline has specific elements tailored for these two classes of targets. As the framework is extended to other modalities, we anticipate adding analogous target-specific controls without compromising the overall automation.

\subsection{From sequence to hits: JAK2 Kinase}

\begin{figure}[htbp]
    \centering
    \includegraphics[width=\textwidth]{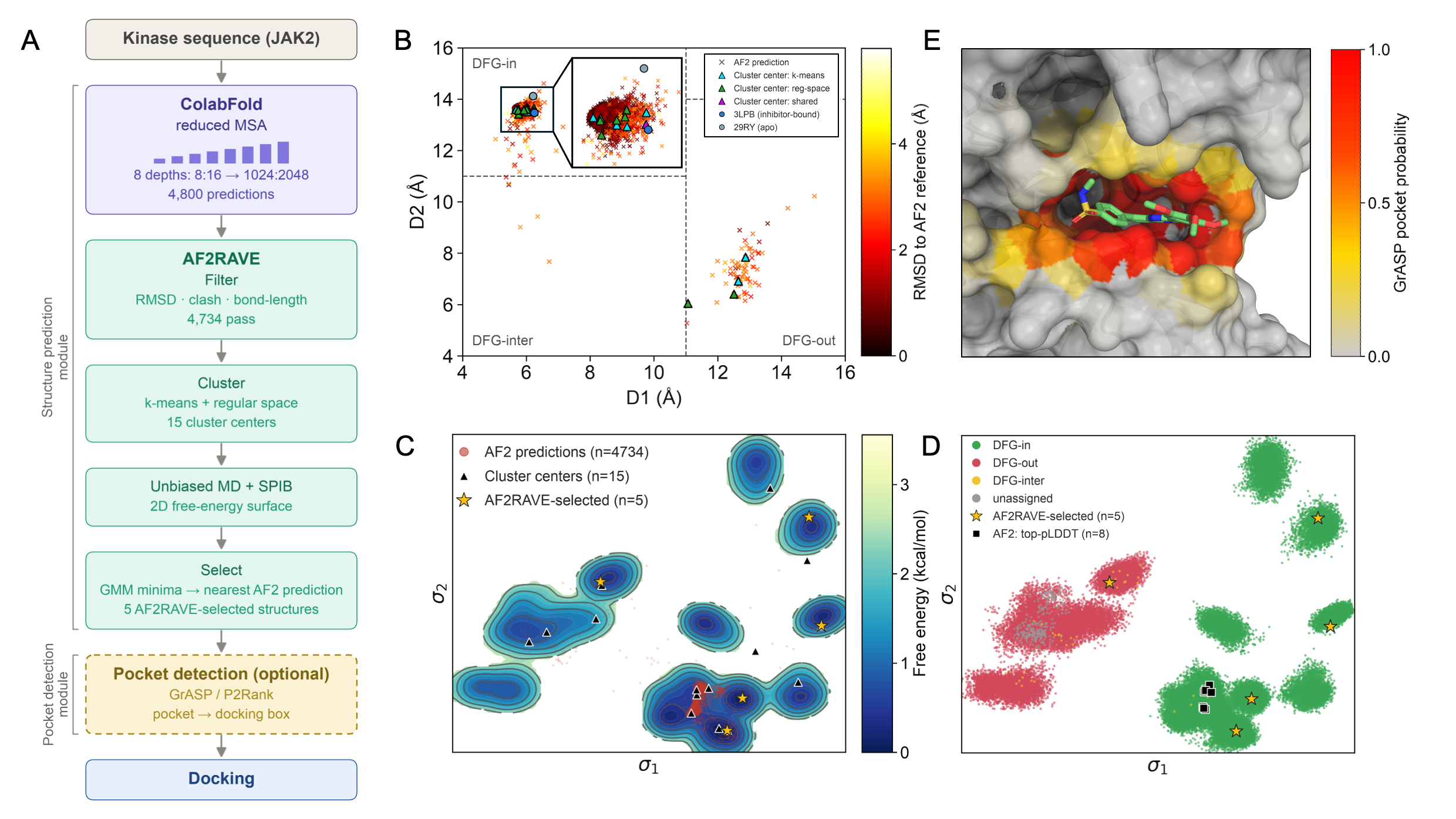}
    \vspace{-12pt}
    \caption{\textbf{From sequence to a docking-ready JAK2 ensemble.}
    \textbf{(A)} Schematic of the conformation-selection pipeline, showing (i)~the JAK2 input sequence,
    (ii)~ColabFold reduced-MSA sampling (8 MSA depths, 4,800 predictions), (iii)~AF2RAVE filtering
    (4,734 pass) and clustering (15 centers), (iv)~unbiased MD with SPIB to learn a free-energy surface,
    and (v)~selection of 5 AF2RAVE conformations for docking. Brackets mark the structure-prediction
    module and the optional pocket-detection module (dashed).
    \textbf{(B)} The 4,734 filtered AF2 structures in Dunbrack D1/D2 space (dashed lines, DFG-state
    boundaries), colored by C$\alpha$-RMSD to the best-pLDDT AF2 reference. Triangles, the 15 cluster
    centers (k-means, teal; regular-space, dark green; shared, purple); circles, crystals 3LPB
    (inhibitor-bound, blue) and 29RY (apo, grey). Inset, DFG-in cluster.
    \textbf{(C)} SPIB free-energy surface ($\sigma_1$, $\sigma_2$; free energy, kcal/mol) with
    the 4,734 AF2 predictions (red), 15 cluster centers (black triangles) and 5 AF2RAVE-selected
    structures (gold stars).
    \textbf{(D)} Same $\sigma_1$/$\sigma_2$ space, MD frames colored by DFG state (in, green; out, red;
    inter, yellow; unassigned, grey); the SPIB coordinates separate DFG-in from DFG-out. Gold stars, the
    5 AF2RAVE-selected structures; black squares, the AF2 top-pLDDT models per MSA depth (n=8).
    \textbf{(E)} GrASP pocket detection on an AF2RAVE-selected conformation; surface colored by
    per-atom pocket probability, with the 3LPB inhibitor (green) superposed on the ATP site. Detection
    accuracy for all five AF2RAVE-selected structures is quantified in SI Fig.~\ref{si:pocket} with GrASP and P2Rank.}
    \label{fig:ensemble}
\end{figure}

\textit{Ensemble generation.} Protein kinases interconvert between active and inactive conformations that differ most visibly at the DFG motif of the activation loop: the active ``DFG-in'' state binds type-I inhibitors, whereas the inactive ``DFG-out'' state is required for type-II inhibitors. Docking a library therefore benefits from receptors that represent both states, yet any single structure---one AlphaFold2 (AF2) model or one crystal---captures only one. Starting from the JAK2 kinase-domain sequence, ANTIOCH-VS generated a diverse set of AF2 models by reduced-multiple-sequence-alignment (reduced-MSA) sampling with ColabFold~\cite{mirdita_colabfold_2022} and kept those passing geometric quality filters (Fig.~\ref{fig:ensemble}A,B). These models already reached both the DFG-in and DFG-out basins, but AF2 predictions carry no thermodynamic weight and remain biased toward the active state~\cite{vani_exploring_2024,gu_empowering_2024,teng_af2rave_2025}, so they cannot be ranked or trusted to represent the relevant metastable conformations.

To place the predictions on a physical footing, ANTIOCH-VS seeds short 50 ns unbiased molecular-dynamics (MD) simulations from representative models and analyzes them with a State Predictive Information Bottleneck (SPIB) model, which learns a low-dimensional free-energy surface separating the kinase states~\cite{wang_spib_2021,mehdi_accelerating_2022}. From this surface it selects a few conformations at the distinct free-energy minima---here five structures spanning both DFG-in and DFG-out---as the docking ensemble (Fig.~\ref{fig:ensemble}C,D). Selecting AF2 models by confidence (pLDDT) instead collapsed onto a single DFG-in cluster (Fig.~\ref{fig:ensemble}D), illustrating why a Boltzmann-weighted ensemble, rather than a confidence-ranked model, is needed to cover the docking-relevant states. With an ensemble in hand, the binding site must be located: for JAK2 the ATP pocket is defined by its reference crystals, and for targets without one the pocket-detection module places the docking box automatically (Fig.~\ref{fig:ensemble}E; SI Fig.~\ref{si:pocket}).

\begin{figure}[htbp]
    \centering
    \includegraphics[width=\textwidth]{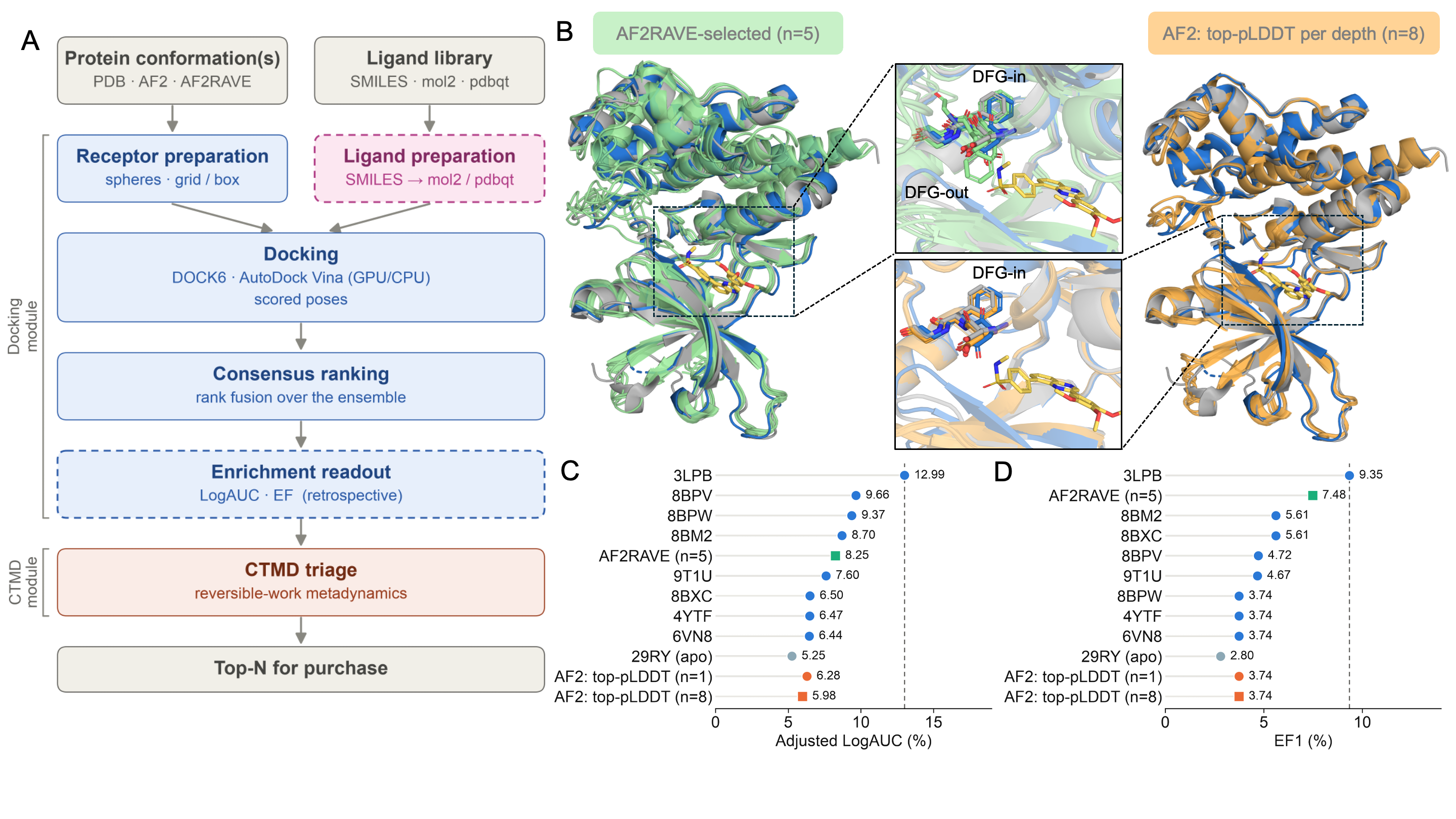}
    \vspace{-35pt}
    \caption{\textbf{JAK2 DUD-E virtual-screening enrichment across receptor sources.}
    Retrospective DUD-E enrichment for the JAK2 kinase domain (107 actives, 6,500 property-matched
    decoys), each receptor docked with DOCK6.
    \textbf{(A)} Schematic of the docking pipeline, showing (i)~receptor and ligand inputs (protein
    conformations from PDB / AF2 / AF2RAVE; a ligand library), (ii)~receptor preparation (spheres,
    grid/box) and optional ligand preparation (dashed), (iii)~docking (DOCK6 or AutoDock Vina),
    (iv)~consensus ranking by reciprocal-rank fusion (RRF) over the ensemble, and (v)~retrospective
    enrichment readout (LogAUC, EF). Brackets mark the docking module and the CTMD module
    (reversible-work metadynamics triage; Fig.~\ref{fig:ctmd}), which together feed the top-$N$ compounds
    for purchase.
    \textbf{(B)} The 5 AF2RAVE-selected (green, left) and 8 AF2 top-pLDDT (orange, right) relaxed
    models, each C$\alpha$-superposed on the inhibitor-bound crystal 3LPB (blue cartoon; NVB inhibitor,
    gold sticks); the apo crystal 29RY is grey. Insets, zoom on the ATP site with each model's DFG motif
    as sticks.
    \textbf{(C)} Adjusted LogAUC and \textbf{(D)} EF1 (actives recovered within the top 1\% of decoys)
    per receptor, ranked high to low; dashed line, the 3LPB reference. Receptors: eight inhibitor-bound
    crystals (3LPB, the DUD-E reference, plus 8BPV, 8BPW, 8BM2, 9T1U, 8BXC, 4YTF, 6VN8), one apo crystal
    (29RY), a single AF2 top-pLDDT model (max\_msa = 1024:2048, $n = 1$), the RRF consensus of 8 AF2 top-pLDDT models
    ($n = 8$), and the RRF consensus of 5 AF2RAVE-selected models ($n = 5$); the two AF2
    points are pinned below the ranking. Color, source class (inhibitor-bound crystal, blue; apo,
    grey-blue; AF2, orange; AF2RAVE, green); shape, single structure (circle) vs.\
    consensus/ensemble (square).}
    \label{fig:enrichment}
\end{figure}

\textit{Virtual-screening enrichment.} We next asked how this sequence-derived ensemble performs in an actual screen, against the receptor choices a real campaign would face (Fig.~\ref{fig:enrichment}). We docked the JAK2 DUD-E library with DOCK6~\cite{balius_dock_2024} against a panel spanning that range: inhibitor-bound crystals (the fortunate case, pocket already primed for the ligand class), an apo crystal (whose pocket may be closed), and three prediction-based options that require no experimental structure of the target---a single high-confidence AF2 model, a confidence-ranked AF2 consensus, and the AF2RAVE-selected ensemble. Because the three prediction-based sources are all AF2 structures differing only in how they were chosen, this panel isolates the effect of the selection strategy itself.

\begin{figure}[htbp]
    \centering
    \includegraphics[width=\textwidth]{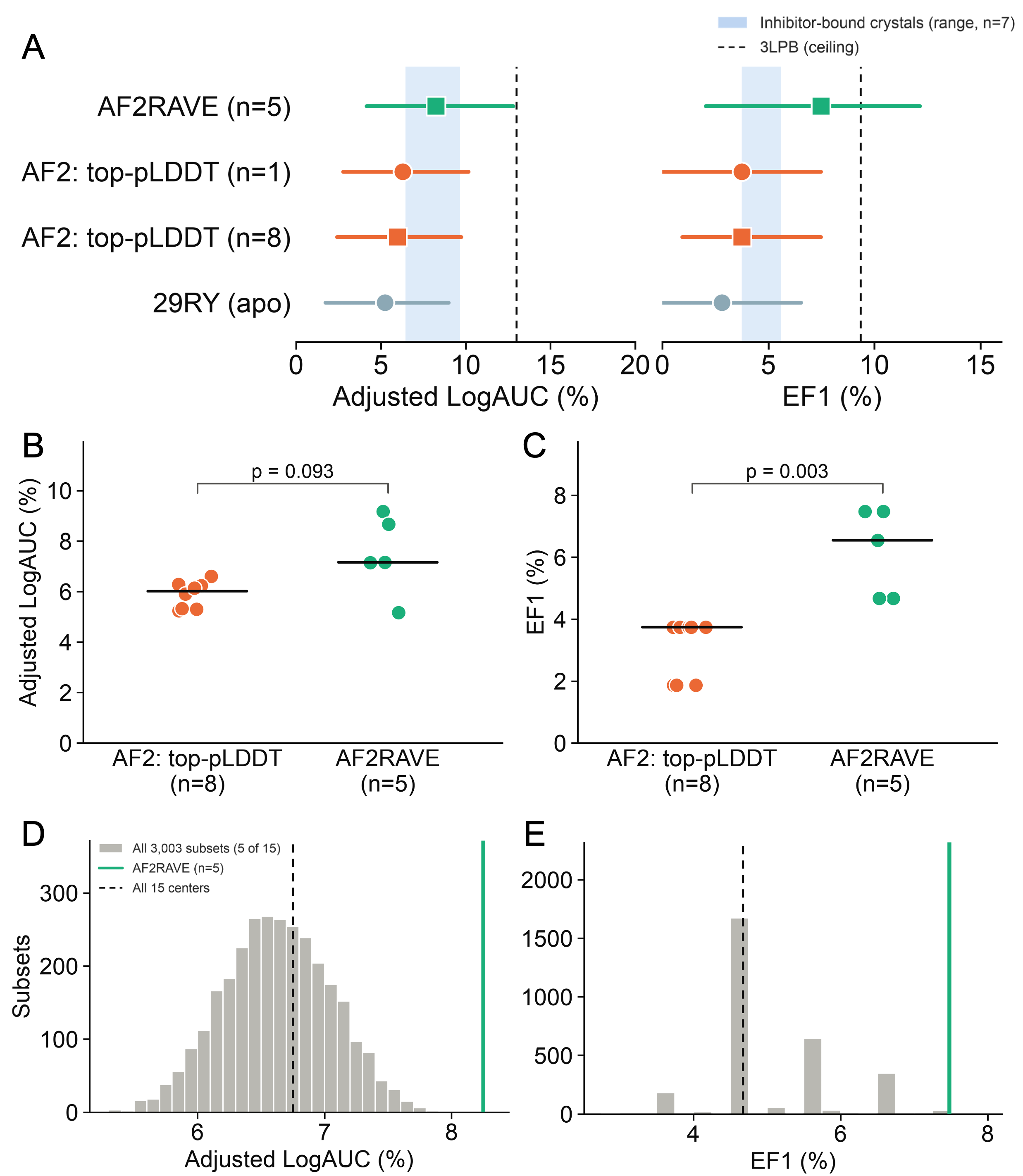}
    \caption{\textbf{AF2RAVE reaches inhibitor-bound-crystal enrichment.}
    \textbf{(A) Docking without an inhibitor-bound crystal.} Adjusted LogAUC (left) and EF1 (right) for
    the apo crystal (29RY), a single AF2 model ($n = 1$), the AF2 consensus ($n = 8$) and the
    AF2RAVE consensus ($n = 5$); points, observed values; bars, 95\% paired-bootstrap confidence
    intervals. Shaded band, the range of the seven inhibitor-bound crystals; dashed line, the 3LPB
    ceiling.
    \textbf{(B, C) Per-model enrichment.} Each point is one receptor model---the 8 AF2 top-pLDDT
    models (orange) versus the 5 AF2RAVE-selected models (green)---for \textbf{(B)} adjusted LogAUC
    and \textbf{(C)} EF1. Horizontal bars, group medians; $p$, two-sided Mann--Whitney (annotated).
    \textbf{(D, E) SPIB selection versus all cluster-center subsets.} Grey histogram, the 3,003
    five-member RRF consensuses drawn from the 15 cluster centers; green line, the AF2RAVE-selected
    consensus ($n = 5$); dashed line, the all-15 cluster center consensus---for \textbf{(D)} adjusted LogAUC and
    \textbf{(E)} EF1. All panels DOCK6.}
    \label{fig:nocrystal}
\end{figure}

Across the panel, the inhibitor-bound crystals enriched most strongly, as expected. Notably, the AF2RAVE ensemble---generated from the JAK2 sequence without using an experimental structure of the target as input---reached into the inhibitor-bound-crystal range on both adjusted logarithmic AUC (LogAUC) and the top-1\% enrichment factor (EF1), and outperformed the alternatives available when no inhibitor-bound crystal exists: the single AF2 model, the AF2 consensus, and the apo crystal, which was the weakest source of all (Fig.~\ref{fig:enrichment}C,D; per-receptor values in SI Table~\ref{si:jak2_enrichment}; ROC curves in SI Fig.~\ref{si:roc}). The confidence-ranked AF2 models sat closest to the inhibitor-bound-crystal backbone but lacked the DFG-out diversity the AF2RAVE ensemble captured (Fig.~\ref{fig:enrichment}B; quantified by C$\alpha$ RMSD to the reference crystals in SI Fig.~\ref{si:ref_rmsd}). These results suggest that ranking conformations by their Boltzmann weight across the metastable free-energy minima, rather than by prediction confidence, is what lets a sequence-derived ensemble approach inhibitor-bound-crystal enrichment. The same qualitative ordering held when the panel was redocked with AutoDock Vina--the AF2RAVE ensemble again enriched above the structure-free alternatives (the single AF2 model, the AF2 consensus, and the apo crystal)--indicating the comparison does not hinge on the docking engine (SI Fig.~\ref{si:vina_enrichment}). We next asked whether this advantage is statistically robust and whether it hinges on the particular five structures SPIB selected (Fig.~\ref{fig:nocrystal}).

\textit{Robustness of the ranking.} A single benchmark split could flatter any one receptor by chance, so we tested whether the ordering holds with a paired bootstrap over the actives and decoys (Methods; full statistics in SI Table~\ref{si:stats}). Restricting to the realistic case with no inhibitor-bound crystal, the AF2RAVE consensus outperformed all three alternatives---the apo crystal and both AF2 baselines---on both metrics, and on EF1 it was statistically indistinguishable from the 3LPB reference crystal (Fig.~\ref{fig:nocrystal}A). The advantage was not merely an averaging effect: compared model by model, the AF2RAVE-selected structures outperformed the AF2 models as a group (Fig.~\ref{fig:nocrystal}B,C).

We also asked whether the result depended on SPIB's particular five-structure choice, or whether any five of the cluster centers would serve as well. Across all 3,003 possible five-member subsets, SPIB's selection ranked at the very top on both metrics, whereas averaging over all fifteen centers was no better than a typical random draw (Fig.~\ref{fig:nocrystal}D,E). These results suggest that the free-energy-based selection, not ensemble averaging, is what drives the enrichment.

\begin{figure}[htbp]
    \centering
    \includegraphics[width=\textwidth]{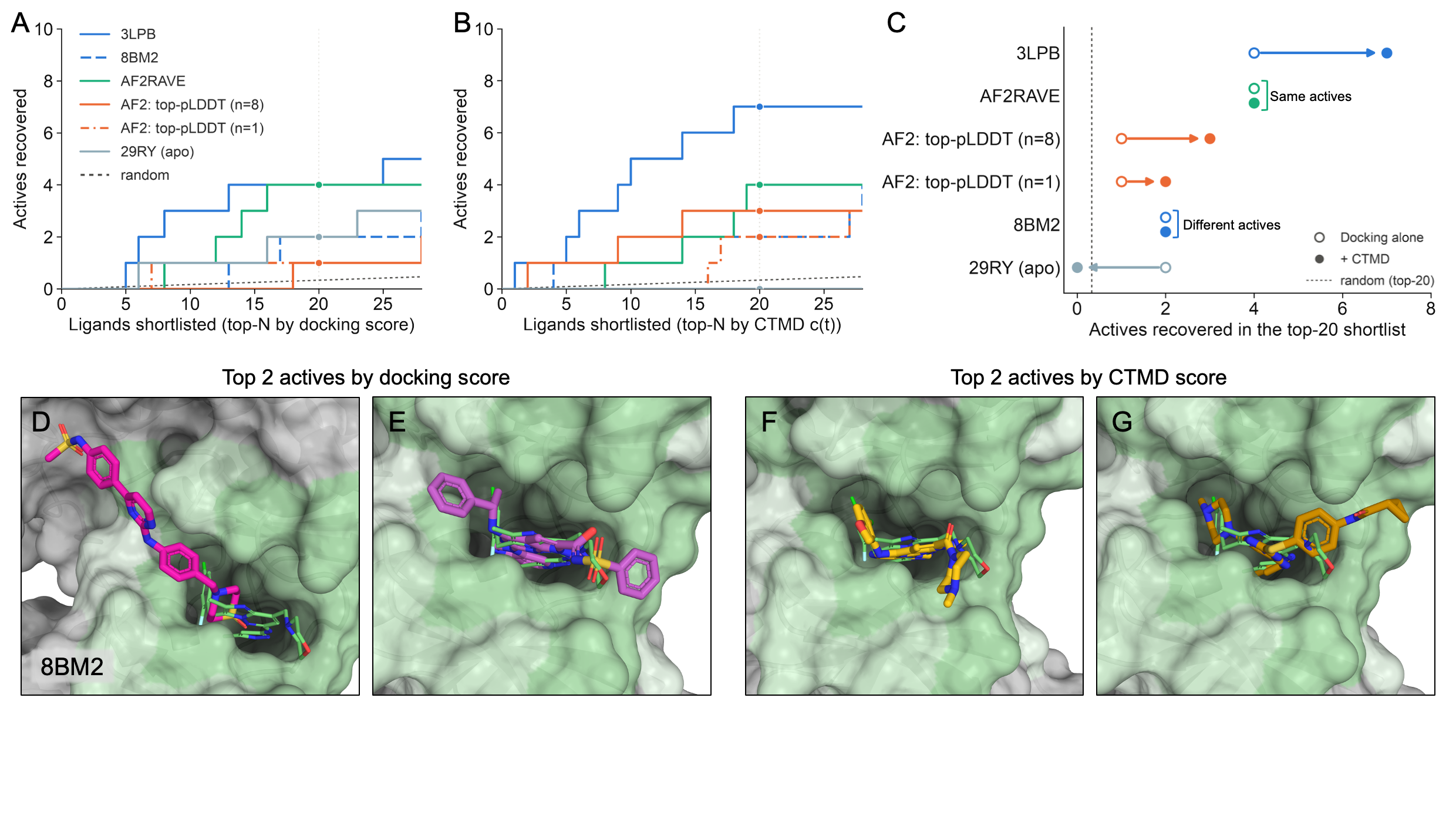}
    \vspace{-45pt}
    \caption{\textbf{CTMD triage of the docked shortlist.}
    Retrospective hit recovery for the JAK2 DUD-E set.
    \textbf{(A, B)} Hit accumulation: cumulative DUD-E actives recovered among the top-$N$ shortlisted
    ligands when each receptor's top-1\% docked poses are ranked by \textbf{(A)} docking score and
    \textbf{(B)} CTMD reversible-work $c(t)$; filled circles mark a 20-ligand budget. Receptors:
    inhibitor-bound crystals 3LPB (solid blue) and 8BM2 (dashed blue), AF2RAVE (green), AF2
    top-pLDDT $n = 8$ (solid orange) and $n = 1$ (dash-dot orange), apo 29RY (grey); dotted line, random
    expectation.
    \textbf{(C)} Actives recovered in the top-20 shortlist for each receptor by docking alone (open) and
    after docking + CTMD (filled), joined by an arrow; dotted line, the random expectation. Labels mark
    receptors whose post-CTMD shortlist holds the same actives versus different actives.
    \textbf{(D--G)} The 8BM2 pocket (grey surface; residues tinted green by proximity to the native
    inhibitor, green sticks) with the top-2 recovered actives by \textbf{(D, E)} docking score (magenta, purple)
    and \textbf{(F, G)} CTMD $c(t)$ (yellow, gold), one active per panel---the ``different actives'' case in
    (C). Of docking's two top-20 actives, one dissociated during the CTMD simulation (D) and the other
    fell from docking rank 13 to CTMD rank 27 (E); CTMD's two, at $c(t)$ ranks 4 and 9 (F, G), were ranked
    63rd and 60th of the 65 triaged poses by docking score.}
    \label{fig:ctmd}
\end{figure}

Having shown that a sequence-derived AF2RAVE ensemble can reach inhibitor-bound-crystal enrichment without an experimental structure of the target, we next asked whether a lightweight physics-based filter could sharpen the resulting shortlist of purchasable compounds (Fig.~\ref{fig:ctmd}).

\textit{CTMD triage of the shortlist.} Docking scores each ligand from a single static pose without explicit water, so a compound can rank well through many contacts in that frozen pose yet not remain bound once solvent and thermal motion are included; rigorous free-energy methods such as FEP~\cite{zwanzig_high-temperature_1954, song_evolution_2020, vogele_conformational_2026} would catch this but are too costly to run across a shortlist. CTMD provides a cheaper physics-based filter: short metadynamics simulations estimate the reversible work $c(t)$ needed to dislodge each docked pose, scoring stably bound poses high and loosely held ones low~\cite{adury_early-enrichment_2026}. Applied as a post-docking triage of each receptor's top-1\% poses, re-ranking a realistic 20-compound shortlist by $c(t)$ recovered more true actives for most receptors---including the inhibitor-bound 3LPB and both AF2 sources---against roughly 0.3 expected from a random selection (Fig.~\ref{fig:ctmd}A--C). The already-strong AF2RAVE shortlist was essentially unchanged, while the apo 29RY shortlist shrank as CTMD rejected its unstable poses.

The inhibitor-bound 8BM2 shows how this triage reshapes a shortlist even when the number of recovered actives is unchanged: docking and CTMD each kept two actives in the top 20, but not the same two (Fig.~\ref{fig:ctmd}D--G). CTMD eliminated a large docking-favored pose perched at the mouth of the pocket that drifted away during simulation, demoted a second whose ring systems extended out of the site, and promoted in their place two deeply buried actives that docking had ranked near the bottom of the shortlist---each overlapping the native 8BM2 inhibitor. An expert might flag such poses by eye, but CTMD does so automatically across the whole shortlist. These results indicate that CTMD re-ranks a docked shortlist by physical binding stability, recovering stably bound actives that a static docking score can overlook.

\subsection{Generalizing from a kinase to a GPCR: ADRB2}

\begin{figure}[htbp]
    \centering
    \includegraphics[width=0.72\textwidth]{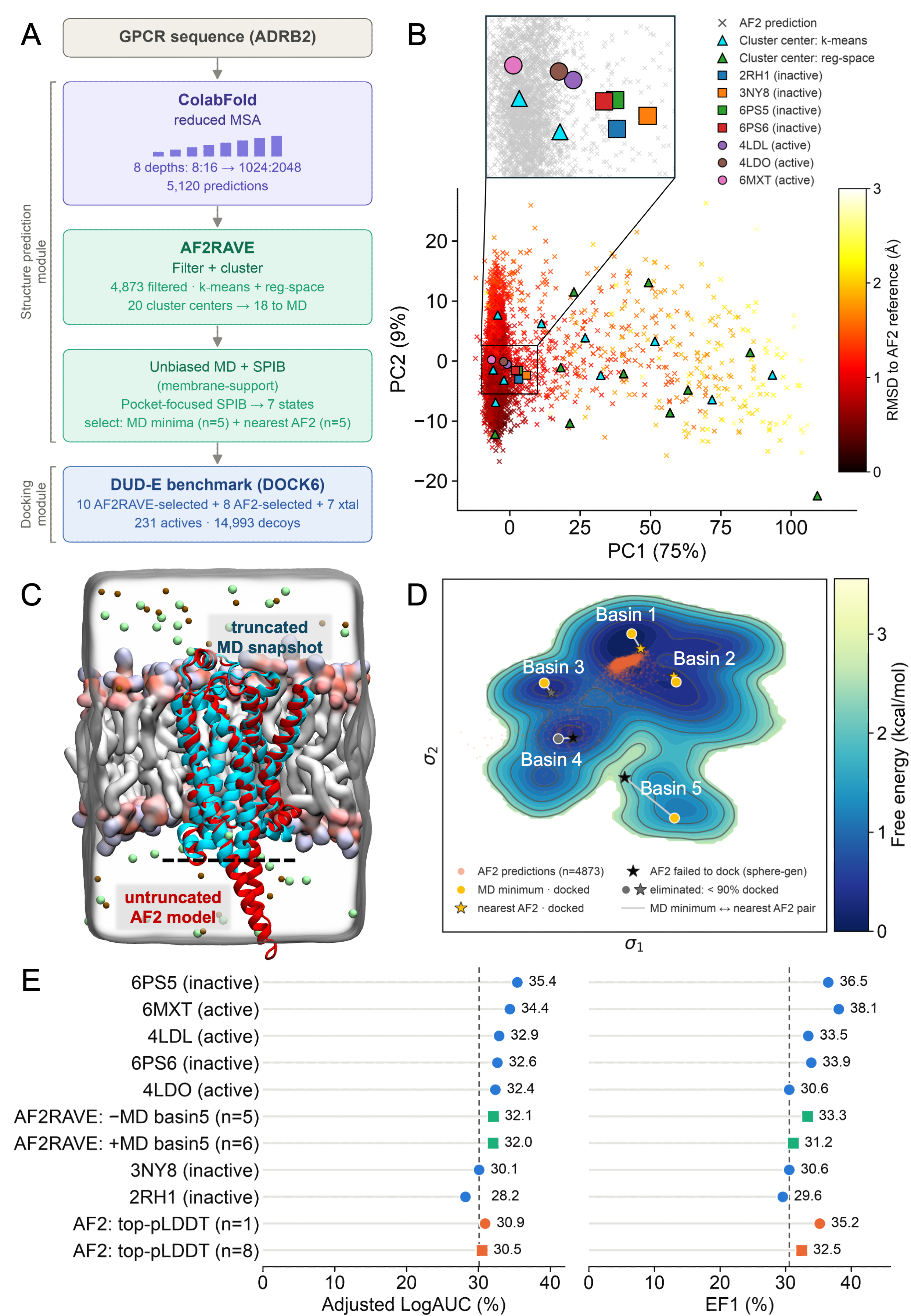}
    \caption{\textbf{Generalizing the sequence-to-hits pipeline from a kinase to a GPCR.}
    The full pipeline applied to ADRB2, benchmarked on its DUD-E set
    (231 actives, 14,993 decoys) with DOCK6.
    \textbf{(A)} Schematic of the pipeline on ADRB2, showing (i)~the ADRB2 input sequence, (ii)~ColabFold
    reduced-MSA sampling (8 depths, 5,120 predictions), (iii)~AF2RAVE filtering (4,873 pass) and
    clustering (20 centers, 18 to MD), (iv)~unbiased membrane MD with pocket-focused SPIB (7
    states), and (v)~selection of 5 MD minima and their 5 nearest AF2 predictions as the
    AF2RAVE-selected conformations ($n = 10$) for docking. Brackets mark the structure-prediction and
    docking modules.
    \textbf{(B)} The 4,873 filtered AF2 structures on the first two principal components of the
    clustering features, colored by C$\alpha$-RMSD to the best-pLDDT AF2 reference. Triangles, the 20
    cluster centers (k-means, teal; regular-space, dark green); the seven ligand-bound ADRB2 crystals
    (one color each) are overlaid and shaped by activation state (squares, inactive: 2RH1, 3NY8, 6PS5,
    6PS6; circles, active: 4LDL, 4LDO, 6MXT). Inset, zoom on the crystals and the cluster centers among
    them.
    \textbf{(C)} A representative AF2RAVE cluster center (rank-1, highest-pLDDT) in an explicit POPC bilayer
    at the end of its unbiased MD simulation. The MD snapshot (cyan; ICL3, residues 200--231,
    truncated before simulation) is superposed on the untruncated AF2 prediction of the same model (red); the dashed line marks approximately where the ICL3 truncation was made. Lipids,
    grey licorice; K$^{+}$ (tan) and Cl$^{-}$ (green) ions as spheres; water omitted.
    \textbf{(D)} Pocket-focused SPIB free-energy surface ($\sigma_1$, $\sigma_2$; free energy, kcal/mol)
    from the unbiased MD simulations, with the AF2 prediction pool ($n = 4{,}873$, orange). The 5 MD minima
    (circles) are each joined to their nearest AF2 prediction (stars); markers are colored by
    docking outcome---gold, kept in the AF2RAVE set; black, a nearest-AF2 that failed DOCK6 sphere
    generation; grey, eliminated for docking $<$90\% of the library.
    \textbf{(E)} Adjusted LogAUC (left) and EF1 (right) per receptor, ranked high to low; dashed line,
    the 3NY8 reference. Receptors: seven ADRB2 crystals (labeled by activation state), two AF2RAVE
    consensus rows (RRF over the kept MD minima and their nearest-AF2 proxies, with the basin-5 MD
    minimum excluded, $n = 5$, or included, $n = 6$), and the AF2 top-pLDDT models ($n = 1$,
    $n = 8$; pinned below the ranking). Color, source class (crystal, blue; AF2RAVE, green; AF2,
    orange); shape, single structure (circle) vs.\ consensus/ensemble (square).}
    \label{fig:adrb2}
\end{figure}

We validated the pipeline first on JAK2 because soluble proteins are comparatively easy to prepare and simulate---an advantage at both the ensemble-generation and CTMD stages. Membrane proteins, however, are among the most important and most challenging drug targets, and G-protein-coupled receptors (GPCRs) are one of their largest therapeutically relevant classes. We therefore added membrane-protein support and applied the same pipeline to ADRB2, chosen because it too has a DUD-E benchmark set and many ligand-bound crystal structures for comparison (Fig.~\ref{fig:adrb2}). 

\textit{Ensemble generation.} From the sequence, ColabFold reduced-MSA sampling produced a filtered ensemble of AF2 models (Fig.~\ref{fig:adrb2}A,B), clustered on general C$\alpha$--C$\alpha$ distance features: whereas prior AF2RAVE work established kinase-specific activation-state features, AF2RAVE had not previously been applied to a GPCR, so we kept the feature set general to favor overall conformational diversity without presupposing the functional coordinates. A physical disulfide-geometry check discarded predictions in which ADRB2's conserved extracellular disulfide could not form. The retained cluster centers were simulated in an explicit lipid bilayer. Because ADRB2's third intracellular loop (ICL3) lies far from the orthosteric pocket and would only enlarge the simulation box, it was truncated before embedding to keep the system compact; the rest of the fold is retained and remains superimposable on the untruncated AF2 prediction of the same model (Fig.~\ref{fig:adrb2}C). A pocket-focused SPIB model trained on the orthosteric-pocket residues then resolved the binding-site conformations into seven metastable states (Fig.~\ref{fig:adrb2}D). Counts, the membrane-system preparation, and SPIB settings are given in Methods and the SI.

\textit{Selection and docking feasibility.} From the pocket-SPIB free-energy surface we selected the five basin minima and docked both each MD snapshot and its nearest AF2 prediction (Fig.~\ref{fig:adrb2}D). Not every candidate yielded a usable pocket: some predicted binding sites were partially closed and failed sphere generation, and others docked too little of the library to be reliable, so these were dropped from the AF2RAVE consensus (per-conformation outcomes in SI Fig.~\ref{si:adrb2_selection} and SI Table~\ref{si:adrb2_rates}). One retained conformation passed this feasibility filter yet enriched poorly in retrospect; computing the consensus with and without it changed the result only slightly, indicating that the AF2RAVE consensus can absorb a single weak member---a useful robustness property when retrospective enrichment is not available to prune members in advance.

\textit{Virtual-screening enrichment.} Benchmarking these sequence-derived receptors against the seven ADRB2 crystals on its DUD-E set (DOCK6; Fig.~\ref{fig:adrb2}E; ROC curves in SI Fig.~\ref{si:roc}), the AF2RAVE consensus again fell within the enrichment band of the crystal structures on both metrics and exceeded the 3NY8 DUD-E reference. It surpassed the AF2 baselines on adjusted LogAUC, though on EF1 a single deep-MSA AF2 model was competitive. The margin between the sequence-derived and crystal receptors was narrower than for the kinase: for ADRB2 even the AF2 baselines enriched close to the crystals (Fig.~\ref{fig:adrb2}E), leaving less headroom for AF2RAVE to improve on them. Even so, the pipeline again produced, from sequence alone, receptors whose enrichment matched experimental crystal structures.

\section{Discussion}\label{sec3}
Together, these results show that the same sequence-to-hits workflow---generating a conformational ensemble, selecting docking-ready receptors, and benchmarking enrichment---transfers across two structurally distinct protein classes, a soluble kinase and a G-protein-coupled receptor, without relying on a target-specific experimental structure. While these two protein classes cover a vast fraction of proteins of therapeutic interest, we will be adding modules to ANTIOCH-VS for other modalities as well.

\textit{What the user must supply.} ANTIOCH-VS is configuration-driven and automates the mechanics of ensemble generation, docking, and triage, but it is deliberately not a fully push-button black box: the quality of the conformational ensemble rests on two choices the user makes up front---which structural features to monitor, and over what range of reduced-MSA depths to sample~\cite{teng_af2rave_2025}. The two case studies illustrate the two regimes. For JAK2, the functionally important motion is well characterized, so we monitored the activation-loop and DFG coordinates that define the kinase active and inactive states, following the feature sets established in prior AF2RAVE kinase studies~\cite{vani_exploring_2024,gu_empowering_2024}; this focuses sampling on the states that matter for inhibitor binding. For ADRB2---a target class not previously studied with AF2RAVE---no such feature set was available, so we used general C$\alpha$--C$\alpha$ distances that favor overall backbone diversity without presupposing the relevant coordinates. The general features still yielded a productive ensemble, though at some cost in focus; where prior knowledge of the functional motion exists, targeting it directly is the more efficient choice.

The feature choice is coupled to the MSA depth. Reduced-MSA sampling trades predictive confidence for conformational diversity: shallower alignments weaken the coevolutionary signal and let AlphaFold2 explore larger, slower conformational changes, whereas deeper alignments converge toward the single most confident fold~\cite{teng_af2rave_2025,delalamo_sampling_2022,wayment-steele_predicting_2024}. Because the appropriate depth depends on the magnitude and timescale of the motion a user wants to capture, we scanned a range of eight depths for both targets rather than committing to one, and let the downstream MD and SPIB steps identify which depths populated the relevant states. We view this modest a priori input---stating the conformational goal through the feature selection and the depth range---as a strength rather than a limitation: it keeps the sampling targeted and the resulting states physically interpretable, in contrast to end-to-end generators that offer less control over which conformations are produced.

\textit{Selecting docking-ready conformations.} The SPIB free-energy surface identifies the metastable states, but the user still chooses which structures to carry into docking, and in what form. Each free-energy minimum can be represented either by the MD snapshot sitting at it or by its nearest AF2 prediction, and the two are not interchangeable. The MD snapshot is a physically equilibrated structure---relaxed in the force field, and within the lipid bilayer for a membrane target---but is a single frame; the nearest AF2 prediction is a clean, reweighted model of the kind AF2RAVE was designed to rank~\cite{vani_alphafold2-rave_2023,teng_af2rave_2025}. For the soluble kinase we docked the nearest-AF2 structures; for the GPCR we docked both the MD minima and their nearest AF2 predictions, which proved worthwhile because the two did not always agree---at two basins the nearest AF2 had a partially closed pocket that failed binding-site generation while the MD snapshot remained dockable. When resources allow, we therefore recommend carrying both representations of each minimum and letting the docking step decide between them.

Not every selected conformation yields a usable binding site, and in a prospective campaign one cannot appeal to retrospective enrichment to tell which will. We found the docking success rate itself to be a practical, enrichment-blind filter: a structure that fails pocket (sphere) generation, or that docks only a small fraction of the library, is signaling an unusable or distorted pocket, so we excluded any conformation that scored fewer than 90\% of the submitted compounds---a rule that removed several GPCR candidates before any enrichment was computed. A complementary safeguard is to screen with a consensus over the retained structures rather than a single model: one retained GPCR conformation still enriched poorly, yet its inclusion barely moved the consensus, showing that ranking over several conformations buffers the occasional weak member. Together these give the user two objective handles that do not require knowing the answer in advance---docking rate as a proxy for pocket quality, and consensus ranking for robustness.

\textit{Reading the DUD-E benchmark outcomes.} Our per-target enrichment under DOCK6 departs sharply from the original DUD-E study~\cite{mysinger_dude_2012}, and in opposite directions for the two targets: on the same receptor structures and decoy sets, we obtain much stronger enrichment than reported for ADRB2 (3NY8 adjusted LogAUC $\approx$30 vs.\ 12, EF1 31 vs.\ 4, ROC AUC $\approx$75 vs.\ 69) but weaker enrichment for JAK2 (3LPB adjusted LogAUC $\approx$13 vs.\ 29, EF1 9 vs.\ 23). The per-receptor Grid-score distributions make the mechanism plain: DOCK6 cleanly separates the ADRB2 actives from decoys at every well-formed pocket, whereas the JAK2 actives and decoys overlap heavily (SI Figs.~\ref{si:adrb2_dockscore},~\ref{si:jak2_dockscore}). Because the difficulty ordering of the two targets inverts relative to the original study---which used the unrelated DOCK3.6 program---this absolute enrichment reflects the particular scoring-function/decoy-set pairing more than an intrinsic property of the target or of our ensembles, a known behaviour of property-matched benchmarks that we discuss further in SI Sec.~\ref{si:benchmark}. We therefore read the benchmarks through within-target, same-decoy comparisons---the AF2RAVE-selected ensembles against the structure-free baselines and the ligand-bound crystals, where decoy-set idiosyncrasies largely cancel---and through the physics-based CTMD triage, rather than through absolute enrichment; the aim here is to show that ANTIOCH-VS runs end to end, not to benchmark docking accuracy.

\textit{Practical challenges.} Two kinds of difficulty recurred. The first is specific to membrane targets: assembling a stable protein--lipid system is more delicate than solvating a soluble domain. In our ADRB2 runs a poorly equilibrated lipid pack could destabilize the system during heating, which we resolved by re-packing the membrane and re-equilibrating rather than by altering the production protocol---a step worth anticipating when extending the pipeline to other membrane proteins. The second difficulty is intrinsic to docking AI-predicted structures: AlphaFold2 models are not guaranteed to respect physical constraints or to present an open binding site. For ADRB2, two predicted conformations placed the conserved extracellular cysteines too far apart to form the native disulfide, and several others had partially collapsed pockets that defeated DOCK6 sphere generation. We handled these with explicit guards---a user-defined disulfide-geometry check that discards physically implausible predictions before simulation, and the docking-rate filter that removes structures whose pockets cannot accommodate the library---rather than assuming every predicted conformation is dockable.

\textit{Ligand preparation.} ANTIOCH-VS accepts ligands either in SMILES (.smi) format (recommended) or directly in the format that is suitable for the docking software chosen (Mol2 for DOCK6, PDBQT for AutoDock Vina). The ``\textbf{antioch pipeline}'' command will automatically perform ligand preparation from the SMILES file provided when it is run after the protein structure prediction module is completed. SMILES are first converted to an internal universal molecular description (UMD) format and then compressed. This format is chemically complete since it stores all bond information, valence, and partial charge calculation. By default, partial atomic charges are calculated using the Open Force Field (OpenFF)~\cite{boothroyd_development_2023} library in python. This UMD format allows for easy mapping of atom identities post docking, allowing easy ligand preparation even for MD simulations. This part was not exercised in this work because the DUD-E libraries were already prepared for DOCK6 in the Mol2 format.

\textit{Containerized, HPC-portable execution.} Most steps in this workflow are computationally expensive and require workstations and HPCs to run in reasonable time. To circumvent the complexities of installing all the required libraries for each step in the process, ANTIOCH-VS ships with a unified Singularity Image File (SIF)~\cite{kurtzer_singularity_2017} which is a self-contained image that contains both our pipeline code as available on GitHub and all required libraries precompiled. If the user intends to use Amber MD as we do, our Amber24 Image, which contains the Amber MD engine, can be plugged in directly~\cite{salomon-ferrer_amber_2013}. This containerization also allows for a more streamlined experience when using our pipeline. The user need only provide a root folder, target sequence, and SMILES file. All required files and paths will be generated instantly when initialized with ``\texttt{antioch init}''. The full file structure generated when using this image is explained in finer detail in Fig.~\ref{si:the_full_pipeline} of the Supplementary Information.

\textit{Limitations and planned extensions.} The present study is retrospective: both campaigns were benchmarked on DUD-E actives and property-matched decoys, and prospective screens remain the decisive test of the pipeline. Several components also have room to grow. The structure-generation front end currently builds ensembles with reduced-MSA ColabFold~\cite{mirdita_colabfold_2022}; we plan to add support for diffusion-based structure predictors, together with ensemble generators tailored to other biomolecular classes such as RNA~\cite{rnaanneal} and intrinsically disordered proteins, to broaden the range of targets the pipeline can address. The molecular-dynamics and CTMD stages currently run through the GPU-accelerated \texttt{pmemd} engine of Amber~\cite{salomon-ferrer_amber_2013}; adding an OpenMM~\cite{eastman_openmm_2017} backend would broaden hardware support and lower the barrier to adoption. CTMD triage was demonstrated here only on the soluble kinase, and extending it to membrane targets---where pose stability must be judged within a bilayer---is a natural next step. We also docked into large, generously defined pockets throughout; for more focused campaigns ANTIOCH-VS supports DOCK6 critical spheres, which require candidate poses to occupy user-specified sub-pocket positions (illustrated for the JAK2 hinge in SI Fig.~\ref{si:critical_spheres}), although we have not yet assessed their effect on enrichment.

\section{Conclusion}\label{sec4}
ANTIOCH-VS is an open-source, modular framework that goes from a protein sequence to a ranked shortlist of candidate binders. We built it to ease two recurring frustrations in virtual screening: stitching many separate tools into one reproducible workflow, and finding a dockable receptor conformation when no ligand-bound crystal is available. The pipeline generates a physically grounded conformational ensemble, docks against it, and triages the top hits---all from a single configuration-driven workflow. We tested it on two very different targets, JAK2 kinase and ADRB2 GPCR, and in both it produced strongly enriched shortlists starting from sequence alone. The code is freely available, and it is built to grow: new structure predictors, docking engines, and post-docking methods can be added as the field evolves.

\newpage
\clearpage

\section{Methods}\label{sec5}
\subsection{Structure Prediction Module}

The structure prediction module provides a configurable interface for generating and analyzing protein structural ensembles for downstream docking workflows. The current implementation integrates AF2RAVE-style ensemble generation and feature selection into the ANTIOCH-VS pipeline, while allowing users to supply either a protein sequence for new structure prediction or an existing folder of predicted structures for post-processing.

For de novo ensemble generation, the module can call a user-provided LocalColabFold installation through a configurable command-line interface. In this case, we use the MSA subsampling approach to ensemble generation. The user specifies the input sequence directly in the configuration file or through a FASTA file, along with ColabFold parameters such as job name, MSA depth, number of seeds, number of recycles, dropout usage, model type, and additional command-line arguments.

When predicted structures are already available, the folding step can be disabled and the module can operate directly on a directory of PDB files. This enables AF2RAVE-style feature selection, clustering, and structure selection to be run independently of structure prediction. The module collects predicted structures, optionally selects a reference structure, applies structural quality filters, ranks structural features, performs regular-space clustering, and writes summary outputs for downstream molecular simulation or docking.

The reference structure can be supplied explicitly by the user. If no reference is provided, ANTIOCH-VS can select the structure with the highest average predicted local distance difference test (pLDDT) score, computed from the pLDDT values stored in the PDB B-factor column. The pLDDT score is averaged at the residue level so that each residue contributes once to the final structure-level score. If pLDDT values are unavailable, the module falls back to using the first loaded structure as the reference.

The module supports several optional structure filters. Structures can be filtered by C-alpha RMSD relative to the reference structure, by minimum non-bonded heavy-atom distance to identify severe steric clashes, and by peptide-bond length statistics to remove structures with abnormal backbone geometry. These filters are only applied when the corresponding cutoffs are provided in the configuration file, allowing some lightweight feature analysis.

For AF2RAVE-style feature selection, users define an atom selection using MDTraj-compatible selection syntax~\cite{mcgibbon_mdtraj_2015}. The module computes pairwise distances between selected atoms across the structural ensemble and ranks these features by coefficient of variation. The top-ranked features represent distances that vary most strongly across the ensemble and can be used as collective variables for clustering or downstream analysis. ANTIOCH-VS writes the ranked features, coefficient-of-variation histogram, filtered structure list, and feature statistics to the output directory.

The selected features are then used for clustering---by either k-means or regular-space clustering---to identify representative structures from the ensemble. The clustering parameters, including distance cutoff, maximum number of centers, batch size, and random seed, are configurable. The resulting cluster-center structures are written to disk and recorded in a summary file. In addition, the module can generate PCA projections of the structural ensemble and mark the selected cluster centers, providing a visual diagnostic of ensemble diversity and representative structure coverage.

Optionally, the structure prediction module can prepare simulation boxes for selected structures. When enabled, the module uses the clustered structures to generate solvated systems with user-defined pH, water padding, ion species, salt concentration, and water model settings, and can also stage the systems for AMBER MD. This provides a bridge between AF2RAVE-style ensemble selection and downstream molecular dynamics workflows, while allowing users to stop at the feature-selection or clustering stage when simulations are not required.

All major steps are controlled through YAML configuration files. The module records a copy of the full configuration in the run log, along with progress messages and output paths, to make each run reproducible. All YAML files are presented to the user immediately after initialization, allowing pre-selection of critical parameters, and not requiring any intervention once the pipeline is launched. The primary outputs include collected structures, filtered structure lists, pLDDT scores, selected feature rankings, cluster-center lists, PCA plots, optional simulation boxes, and a run report summarizing the completed workflow.

\subsection{Receptor and Ligand Preparation for Docking}

ANTIOCH-VS includes a lightweight receptor- and ligand-preparation utility intended to help users quickly move from a protein structure to docking-ready inputs. The user can provide either a local complex structure or a PDB identifier. When a PDB identifier is provided, the script fetches the corresponding structure, extracts the requested receptor chains, and identifies the ligand specified by residue name, chain identifier, or residue number as needed.

For receptor preparation, the utility uses PDBFixer~\cite{eastman_openmm_2017} to repair common issues in the protein structure. Nonstandard residues are replaced where possible, missing heavy atoms are added, and hydrogens are assigned at a user-defined pH. The user can choose whether to retain crystallographic waters and whether to model missing residues. Because automatic loop modeling can be unreliable for difficult cases, the preparation report records what was added and warns users when missing residues may require more careful manual modeling.

The script also supports optional receptor-chain selection. To preserve information about missing residues, receptor preparation is first performed on the receptor-containing complex before the requested receptor chains are written out. This avoids losing missing-residue annotations that may be present in the original PDB file. The final output is a prepared receptor PDB file that can be passed directly to the docking-specific preparation steps.

Ligand extraction is performed from the original complex using the user-specified ligand residue. For DOCK6 workflows, the utility writes a reference ligand Mol2 file without hydrogens for binding-site sphere selection and can also write a prepared ligand Mol2 file for docking. For Vina workflows, the extracted or user-provided ligand can be converted to PDBQT format for docking. The preparation report includes basic ligand properties such as atom count and formal charge, allowing the user to decide whether more careful ligand preparation is needed.

This preparation utility is intentionally lightweight. It is designed for rapid testing, benchmark setup, and simple redocking workflows rather than production-grade receptor or ligand curation. For careful virtual screening studies, users may still provide externally prepared receptors and ligands generated with their preferred structure-preparation tools. ANTIOCH-VS therefore supports both quick automated setup and fully user-controlled inputs.

\subsection{Docking Module: DOCK6 Pipeline}

The DOCK6 pipeline provides an automated workflow for receptor preparation and flexible ligand docking with UCSF DOCK6. It connects receptor preparation, molecular surface generation, sphere selection, grid generation, and flexible docking through a configuration-driven interface. The user provides a prepared receptor structure, a reference ligand defining the binding site, and one or more ligands to dock in Mol2 format.

Receptor preparation is handled through an integrated UCSF Chimera~\cite{pettersen_ucsf_2004} workflow. Starting from a receptor PDB file, the Chimera pipeline can add hydrogens, assign charges, write a charged receptor Mol2 file, and generate a molecular surface in Dot Molecular Surface (DMS) format. The DMS surface is used by DOCK6 for sphere generation, while the charged receptor Mol2 file is used during grid construction. Chimera can be invoked through a local installation, a micromamba environment, or a Singularity container, allowing the same workflow to be used across workstations and HPC systems.

After receptor preparation, the DOCK6 pipeline generates spheres with \texttt{sphgen} from the receptor DMS surface. The resulting sphere file represents possible ligand-binding regions on the receptor surface. A reference ligand Mol2 file is then used with \texttt{sphere\_selector} to retain only spheres near the known or user-defined binding site. This allows the docking region to be defined from a crystallographic ligand, a guide ligand generated by the receptor-preparation module, or another ligand-like molecule placed in the desired pocket.

The selected spheres are used to construct a docking box with \texttt{showbox}. The user can control the sphere-selection radius and the box margin through the configuration file. This makes the search region adjustable for small redocking tests, focused virtual screens, or broader pocket exploration. The box and receptor Mol2 file are then passed to \texttt{grid}, which precomputes receptor interaction grids for subsequent ligand scoring. The pipeline writes all intermediate files, including sphere files, box files, grid input files, and grid outputs, into organized subdirectories.

Flexible docking is performed with \texttt{dock6}. The pipeline writes a DOCK6 flexible-docking input file using conservative defaults, while allowing users to override individual DOCK6 parameters through the YAML configuration. The default mode uses flexible ligand growth, grid-based scoring, ligand minimization, anchor minimization, and flexible-growth minimization. Users can adjust sampling parameters such as the maximum number of orientations, number of final scored poses, number of preclustered conformers, simplex minimization settings, and other DOCK6 options without modifying the source code.

The pipeline also supports pose RMSD calculation during redocking. When enabled, DOCK6 reports heavy-atom RMSD values for docked poses relative to the ligand input used as the RMSD reference. This is useful for validating whether the workflow can recover a known crystallographic ligand pose before running larger ligand libraries. The included DUD-E kinase examples (ABL1 and CDK2) demonstrate crystallographic pose recovery and the effect of orientation- and conformer-sampling settings.

Both native and containerized DOCK6 execution are supported. In native mode, the user specifies the path to a local DOCK6 installation containing the \texttt{bin} and \texttt{parameters} directories. In Singularity mode, the user provides a DOCK6 \texttt{.sif} image, optional bind paths, and, if needed, custom command names for the DOCK6 executables inside the container. The pipeline automatically binds the relevant input and output directories and runs \texttt{sphgen}, \texttt{sphere\_selector}, \texttt{showbox}, \texttt{grid}, and \texttt{dock6} through the container. This design makes the workflow portable across HPC environments where software availability and installation paths differ.

All DOCK6 pipeline steps are controlled by YAML configuration files and are written to structured output directories. The run log records the full configuration snapshot, executed steps, standard output, and standard error from each external program. This makes it possible to inspect failures at the level of individual DOCK6 stages, repeat runs with modified sampling parameters, and compare receptor or ligand preparation choices across benchmark systems.

\subsection{Docking Module: AutoDock Vina and Vina-GPU Pipelines}

The docking module also provides automated pipelines for running AutoDock Vina and Vina-GPU docking. These workflows are intended for high-throughput virtual screening in which large ligand libraries are docked against one or more prepared receptor structures. Both CPU and GPU execution modes are supported through separate configuration files, while preserving the same general input structure and output organization.

The Vina CPU pipeline starts from one or more receptor PDB files and a directory of ligands in PDBQT format. For each receptor, the pipeline prepares the receptor for docking, writes a Vina configuration file, and runs batch docking against the ligand set. The docking box is specified through user-defined center coordinates and box dimensions in the YAML configuration file. This allows the user to define the search region from a known ligand, a known binding pocket, or coordinates obtained from the receptor-preparation module.

The Vina-GPU pipeline is designed for accelerated screening of large ligand libraries on GPU-enabled systems. It uses the same conceptual inputs as the CPU pipeline: prepared receptor structures, ligand PDBQT files, docking-box coordinates, and Vina-style docking parameters. The pipeline organizes GPU docking runs into structured output directories and writes run logs for each screen. This enables the same receptor and ligand preparation workflow to be used for smaller CPU tests and larger GPU screens.

Both Vina pipelines support execution through native installations or Singularity containers. When a Singularity image is provided, the pipeline detects the containerized execution mode and calls the appropriate internal Vina command. Required host directories can be bind-mounted through the configuration file, which allows the workflow to run on HPC systems where Vina or Vina-GPU are distributed as container images. This avoids hard-coding cluster-specific paths and makes the same configuration pattern portable across computing environments.

The module writes Vina configuration files automatically from the user-provided YAML settings. These include the receptor path, ligand input path, docking center, box dimensions, exhaustiveness or search-depth parameters where applicable, output location, and logging paths. By generating these configuration files programmatically, ANTIOCH-VS reduces manual editing and makes each docking run easier to reproduce.

The Vina workflows are compatible with receptor and ligand files generated by the receptor-preparation module. For Vina-style docking, ligands are prepared in PDBQT format, and receptors are converted to the required docking format before screening. The receptor-preparation module can also report the center of mass of a reference ligand, which provides a convenient starting point for defining the Vina docking box. This is especially useful when the user begins from a PDB structure containing a bound ligand and wants to set up a focused docking calculation around the known binding site.

After docking, the Vina pipelines collect docking outputs and logs in organized output directories. Docking results can be further processed using the extraction utilities included in ANTIOCH-VS. These utilities summarize docking scores across ligand files, identify the top-ranked compounds, and export selected poses for visualization or downstream analysis. In this way, the Vina pipelines provide both the docking execution layer and the first-pass result organization needed for virtual screening campaigns.

Note that the Vina module separates CPU and GPU execution because current Vina implementations differ in supported features. Vina-GPU 2.1 provides substantial acceleration for large-scale screening, but it was developed from the older AutoDock Vina 1.1.2 codebase. Therefore, newer Vina 1.2.x features, including macrocycle docking support, are not available in Vina-GPU 2.1. For macrocyclic ligands or workflows requiring recent AutoDock Vina features, users can run the Vina CPU pipeline with an up-to-date Vina executable. For large nonmacrocyclic libraries where throughput is the priority, the Vina-GPU pipeline provides the accelerated option.

\subsection{CTMD Module: Post-Docking Stability Triage}

The CTMD module re-ranks docked poses by physical binding stability using $c(t)$-based reversible-work metadynamics~\cite{adury_early-enrichment_2026}. For a selected receptor--ligand complex, the module automates AMBER system preparation~\cite{case_ambertools_2023}: ligand partial charges are assigned with \texttt{antechamber} (AM1-BCC~\cite{jakalian_am1bcc_2002}) and any missing parameters with \texttt{parmchk2} under the GAFF2 force field~\cite{wang_gaff_2004}, the receptor is taken from the docking-prepared structure, and \texttt{tleap} builds the complex topology, solvates it in explicit water, and neutralizes the net charge. Each solvated complex is then energy-minimized, heated, and equilibrated with positional restraints on the receptor C$\alpha$ atoms before production.

Pose stability is probed by well-tempered metadynamics~\cite{barducci_well-tempered_2008} in PLUMED~\cite{tribello_plumed_2014}. A bias is applied along a protein-aligned ligand-RMSD coordinate (relative to the docked pose) to drive the ligand out of its docked pose, and several short metadynamics replicas (each on the order of 10~ns) are run per complex with the GPU-accelerated \texttt{pmemd} engine of Amber~\cite{salomon-ferrer_amber_2013} at constant temperature and pressure. For each replica, the reversible work accumulated up to the point at which the ligand commits to leaving the pocket is read from the time-dependent metadynamics reweighting factor $c(t)$~\cite{tiwary_time-independent_2015}: a pose that resists displacement accumulates large reversible work, whereas one that dissociates quickly accumulates little. The per-ligand CTMD score is a conservative bootstrap estimate of this reversible work---the mean, over resampled subsets of the replicas, of the lowest $c(t)$ within each subset---and ligands are ranked by this score, higher $c(t)$ indicating a more stable pose. Complexes whose simulations terminate early because the ligand leaves the pocket, or that yield too few valid replicas, receive no $c(t)$ and are treated as unstable and dropped from the triaged list; this is the intended behavior, since early ligand departure is itself evidence that the docked pose is not stably bound.

\subsection{Containerized Execution with Singularity}

ANTIOCH-VS supports containerized execution through Singularity, enabling the same workflow to run on local workstations and HPC systems where software installations are managed through container images. This is particularly important for docking workflows, where programs such as UCSF Chimera, DOCK6, AutoDock Vina, and Vina-GPU may have different dependency requirements or may not be available as native installations on every system.

For each supported external program, the user can choose between native execution and Singularity-based execution in the YAML configuration file. In native mode, ANTIOCH-VS calls the executable directly from the host environment. In Singularity mode, the user provides the path to a \texttt{.sif} image and, when necessary, the command names or executable paths inside the container. ANTIOCH-VS then wraps the external program call with \texttt{singularity exec}, while preserving the same pipeline logic and output structure.

The pipeline automatically bind-mounts the relevant input and output directories needed for each run. Users can also provide additional bind paths in the configuration file when working on HPC systems with separate project, scratch, or software filesystems. This allows containerized tools to access receptor files, ligand libraries, output directories, and shared software resources without hard-coding cluster-specific paths into the pipeline.

Singularity support is implemented across the docking workflows. The Chimera receptor-preparation pipeline can run through a Chimera container to add hydrogens, assign charges, generate receptor Mol2 files, and write DMS molecular surfaces. The DOCK6 pipeline can run \texttt{sphgen}, \texttt{sphere\_selector}, \texttt{showbox}, \texttt{grid}, and \texttt{dock6} from a DOCK6 container. The Vina and Vina-GPU pipelines can similarly detect containerized execution and call the appropriate internal Vina command inside the image.

This design makes ANTIOCH-VS portable across different computational environments. A user can test a workflow locally with native binaries, then move the same YAML-driven workflow to an HPC cluster by specifying container images and bind paths. On some HPC systems, users may need to load the Singularity or Apptainer module before running ANTIOCH-VS, for example with \texttt{module load singularity}. Once the container runtime is available, ANTIOCH-VS handles the construction of the external commands and records the executed configuration in the run log for reproducibility.

\subsection{Configuration and Reproducibility}

All ANTIOCH-VS workflows are controlled through YAML configuration files. Each module reads user-defined paths, execution settings, and method-specific parameters from a configuration file, allowing workflows to be rerun or transferred across computing environments with minimal manual modification. During execution, ANTIOCH-VS records the full configuration snapshot in the run log, along with progress messages, external command output, and error messages. This makes each run easier to audit, debug, and reproduce.

The pipeline is initialized only once per problem statement with \texttt{antioch init} where it writes template YAML files with reasonable defaults. The user can edit them before launching the full pipeline. Outputs are written into structured module-specific directories. Intermediate files, generated input files, docking logs, scored poses, selected structures, and summary files are preserved unless the user explicitly overwrites a previous run. This organization allows users to inspect individual stages of the workflow, repeat failed steps, and compare alternative receptor, ligand, or docking parameter choices.

\subsection{Benchmark Targets, Datasets, and Receptor Panels}

For JAK2 we used the kinase domain and its DUD-E benchmark set~\cite{mysinger_dude_2012} of 107 actives and 6{,}500 property-matched decoys. The receptor panel comprised eight inhibitor-bound crystal structures---3LPB (the DUD-E reference), 8BPV, 8BPW, 8BM2, 9T1U, 8BXC, 4YTF, and 6VN8---one apo crystal (29RY), and three sequence-derived receptor sources: a single AlphaFold2 (AF2) model (the rank-1 top-pLDDT model at the deepest reduced-MSA depth, \texttt{max\_msa} 1024:2048; $n=1$), the reciprocal-rank-fusion (RRF) consensus of the eight top-pLDDT models, one per MSA depth ($n=8$), and the RRF consensus of the five AF2RAVE-selected structures ($n=5$). For ADRB2 we used its DUD-E set (231 actives and 14{,}993 decoys) and a panel of seven ligand-bound crystals spanning inactive and active states (Fig.~\ref{fig:adrb2}); its ensemble-generation specifics are given with the ADRB2 results.

\subsection{Conformational-Ensemble Generation (JAK2)}

Starting from the JAK2 kinase-domain sequence, the structure-prediction module ran ColabFold~\cite{mirdita_colabfold_2022} with reduced-MSA subsampling across eight MSA depths, generating 4{,}800 candidate models. These were passed through the module's geometric filters---C$\alpha$-RMSD to the highest-pLDDT reference model, a minimum nonbonded heavy-atom distance to flag steric clashes, and peptide-bond-length checks---leaving 4{,}734 models. For feature selection, pairwise C$\alpha$--C$\alpha$ distances were computed over the activation loop and the adjacent $\alpha$C-helix/$\beta$3 region, together with sidechain atoms of the DFG motif, the $\beta$3 lysine, and the $\alpha$C glutamate---the atoms defining the Dunbrack D1/D2 kinase-state coordinates (exact MDTraj selection in SI Sec.~\ref{si:methods})---and ranked by coefficient of variation. The top-ranked features were clustered by two complementary methods, k-means (density-based) and regular-space (diversity-promoting), taking eight centers from each (one shared) for 15 representative cluster centers. A 50~ns unbiased MD simulation was seeded from each center, and the pooled trajectories were used to train a State Predictive Information Bottleneck (SPIB) model~\cite{wang_spib_2021,mehdi_accelerating_2022} (time-lag and network hyperparameters in SI Sec.~\ref{si:methods}), yielding a two-dimensional free-energy surface that separates the DFG-in and DFG-out states. A Gaussian mixture model was fit to this surface, and the five free-energy minima were mapped to their nearest AF2 predictions to form the AF2RAVE docking ensemble. The $n=8$ AF2 baseline took the rank-1 top-pLDDT model from each MSA depth, and the $n=1$ baseline took the deepest-MSA model.

\subsection{Conformational-Ensemble Generation (ADRB2)}

Starting from the ADRB2 sequence, ColabFold~\cite{mirdita_colabfold_2022} reduced-MSA sampling across eight MSA depths generated 5{,}120 candidate models, of which 4{,}873 passed the geometric filters. In contrast to the kinase, for which prior AF2RAVE studies had established activation-state-specific features, feature selection here used general pairwise C$\alpha$--C$\alpha$ distances across the receptor (MDTraj selection \texttt{name CA}), ranked by coefficient of variation, so as not to presuppose the receptor's functional coordinates. The top-ranked features were clustered by k-means and regular-space clustering into 20 centers (ten from each method); two centers were then removed by a disulfide-bond check---discarding predicted conformations in which ADRB2's conserved extracellular (ECL2) disulfide could not form---leaving 18 centers for molecular dynamics. Each center was embedded in an explicit lipid bilayer prepared with Packmol-Memgen~\cite{schott-verdugo_packmol-memgen_2019}, chosen for compatibility with the \texttt{pmemd} engine of Amber underlying CTMD, and propagated with unbiased MD. A pocket-focused SPIB model~\cite{wang_spib_2021,mehdi_accelerating_2022} trained on the residues lining the 3NY8 orthosteric pocket resolved the binding-site conformations into seven metastable states. The five free-energy minima of the resulting surface, together with their five geometrically nearest AF2 predictions, formed the ten candidate receptors carried to docking (exact feature list, membrane-preparation parameters, and SPIB hyperparameters in SI Sec.~\ref{si:methods}).

\subsection{Selecting Docking Conformations from the SPIB Surface}
For both targets the docking conformations were chosen from the two-dimensional SPIB free-energy surface by the same Gaussian-mixture procedure (SI Fig.~\ref{si:spib_gmm}). A Gaussian mixture model (\texttt{scikit-learn}~\cite{pedregosa_scikit-learn_2011} \texttt{GaussianMixture}, \texttt{random\_state}=9) with five components---the target number of docking conformations---was fit to the unbiased-MD frames in SPIB space, partitioning the sampled density into five regions. Within each component, the single frame of lowest free energy (estimated as $-k_{\mathrm B}T\ln\rho$ from the sampled SPIB-space density) was taken as that component's representative free-energy minimum, yielding five minima per target. The component count sets how many conformations are carried forward and is distinct from the number of SPIB metastable states (seven for the ADRB2 pocket-SPIB model); fitting five components over the surface merges nearby basins so that the five selected minima are spread across the populated landscape.

For JAK2 the five minima were mapped to their geometrically nearest AF2 predictions, and those five structures form the AF2RAVE-selected consensus ($n=5$) used throughout (SI Fig.~\ref{si:spib_gmm}, left). For ADRB2 the five MD-minimum snapshots (md1--md5) and their five nearest AF2 predictions were both carried to docking as the ten candidate receptors, after which the docking-rate filter and sphere-generation check reduced them to the AF2RAVE-selected consensus (SI Fig.~\ref{si:spib_gmm}, right; Table~\ref{si:adrb2_rates}). The JAK2 SPIB surface resolves into well-separated basins, whereas the ADRB2 pocket-SPIB surface is more collapsed, with adjacent overlapping components, reflecting the narrower conformational spread of the orthosteric pocket.

\subsection{Docking Setup (JAK2)}

JAK2 receptors were docked against the DUD-E library with DOCK6~\cite{balius_dock_2024} using the pipeline's default flexible-docking parameters, which were identical across all receptors of both targets (full sphere, grid, and flexible-docking parameter listing in SI Table~\ref{si:docking_params}). The docking box was defined from the ATP pocket with the DOCK6 sphere-selection and grid workflow, using the 3LPB reference ligand as the binding-site guide. Where a pocket is not known in advance, the pocket-detection module (GrASP~\cite{smith_grasp_2024}; P2Rank~\cite{krivak_p2rank_2018}) recovered the orthosteric ATP site on the five AF2RAVE structures (SI Fig.~\ref{si:pocket}). For the AF2 ($n=8$) and AF2RAVE ($n=5$) ensembles, per-receptor rankings were combined by unweighted RRF before computing enrichment. As a functional check that both docking engines run end to end in the pipeline, the co-crystallized 3LPB ligand was redocked with DOCK6 and AutoDock Vina; both recovered the crystallographic pose within 2 Å (SI Fig.~\ref{si:redocking}).

\subsection{Docking Setup (ADRB2)}

ADRB2 receptors were docked against the DUD-E library with DOCK6~\cite{balius_dock_2024}, with the docking box defined from the 3NY8 orthosteric pocket. Both the MD free-energy-minimum snapshots and their nearest AF2 predictions were docked (ten candidates in all), and each was screened by its docking rate against the submitted library (447 active and 15{,}255 decoy states; see the enrichment subsection below). Candidates that failed DOCK6 sphere generation, or that docked fewer than 90\% of the submitted states, were excluded from the AF2RAVE-selected consensus (per-conformation outcomes in SI Table~\ref{si:adrb2_rates} and SI Fig.~\ref{si:adrb2_selection}). The retained conformations were combined by unweighted RRF; because one retained MD minimum (basin~5) enriched poorly in retrospect, the consensus was reported both including it ($n=6$) and excluding it ($n=5$).

\subsection{Post-Docking CTMD Triage of the Benchmark Shortlists}

Each protein--ligand pair was prepared by the same procedure used for the initial MD simulations, except that the ligands were parameterized with \texttt{antechamber} using the GAFF2 force field, with partial charges from the AM1-BCC method. The simulation protocol was otherwise identical, except that the final equilibration MD was only 5~ns. For the ligands that remained, 10 metadynamics trajectories were launched from the final frame of this 5~ns MD. The biased collective variable (CV) was the protein-aligned ligand RMSD, with the protein aligned to the start of the unbiased MD as reference.

Each well-tempered metadynamics~\cite{barducci_well-tempered_2008} run lasted at most 5~ns, or until the ligand exceeded its predefined RMSD upper bound, set to 5~\AA{} here. A hill height of 1.75~kJ/mol was deposited every 1~ps with a bias factor of 10. Final CTMD analysis used a commit time of 1~ns.

Because each metadynamics run is far more costly than docking, CTMD was applied in this study only to the top 1\% of the consensus ranking. Hit recovery was then evaluated by re-ranking this triaged shortlist by $c(t)$ and counting the DUD-E actives retained within a fixed purchase budget of 20 compounds (Fig.~\ref{fig:ctmd}).

\subsection{Enrichment Metrics and Statistical Analysis}

Retrospective enrichment was quantified on the DUD-E benchmark sets (JAK2: 107 actives and 6,500 property-matched decoys; ADRB2: 231 actives and 14,993 decoys) using three metrics computed with tie-aware rank statistics: the area under the ROC curve, the enrichment factor in the top 1\% of ranked decoys (EF1), and the DUD-E adjusted logarithmic AUC (adjusted LogAUC) integrated from 0.1\% decoys (illustrated in SI Fig.~\ref{si:logauc_explainer}). The number of decoys that yield a valid docking score varies slightly between receptors and between docking engines because a small fraction of compounds fail pose generation on a given structure; metrics for each receptor were computed on the compounds it successfully scored. For receptor ensembles, per-receptor rankings were combined by unweighted reciprocal-rank fusion (RRF) before computing enrichment.

For ADRB2, the 231 active and 14{,}993 decoy unique DUD-E ligands were expanded to 447 active and 15{,}255 decoy protomer/tautomer states (15{,}702 states) for docking. Each AF2RAVE candidate receptor's docking rate---the fraction of these submitted states that produced a valid pose---was used to filter the selection: conformations that docked fewer than 90\% of the submitted states were excluded from the AF2RAVE-selected consensus, and conformations that failed DOCK6 sphere generation outright contributed no docking at all (per-conformation rates, outcomes, and single-receptor enrichment in SI Table~\ref{si:adrb2_rates} and SI Fig.~\ref{si:adrb2_selection}).

For the no-inhibitor-crystal comparison (Fig.~\ref{fig:nocrystal}A and Table~\ref{si:stats}), confidence intervals were obtained by a stratified paired bootstrap. Only ligands scored by every receptor of a given engine were retained (for DOCK6, 107 actives and 6,327 decoys), and point estimates were recomputed on this shared set. Actives and decoys were then resampled with replacement separately (2,000 iterations; NumPy \texttt{default\_rng}, seed 20261006), and the same resampled ligand set was applied to every receptor so that between-receptor differences are paired. Reported 95\% intervals are percentile intervals (2.5th--97.5th). For each receptor, the one-sided significance of the AF2RAVE consensus advantage is the fraction of resamples in which the AF2RAVE consensus did not exceed that receptor, $p = \mathrm{mean}(\Delta \le 0)$ with $\Delta$ the paired difference (AF2RAVE consensus minus receptor). Per-model groups (the five AF2RAVE-selected structures versus the eight AF2 top-pLDDT models) were compared with a two-sided Mann--Whitney $U$ test. To test the SPIB selection against chance, all $\binom{15}{5} = 3{,}003$ five-member subsets of the cluster centers were enumerated and each fused into an RRF consensus, and the SPIB-selected consensus was ranked within the resulting distribution.

\backmatter

\bmhead{Supplementary information}

\bmhead{Acknowledgements}

This work was supported by NIH/NIGMS under Award No. R35GM142719. We thank UM-IHC Beacon, UMD Zaratan and NSF ACCESS (Project CHE180027P) for computational resources. P.T. is an investigator at the University of Maryland-Institute for Health Computing, which is supported by funding from Montgomery County, Maryland, and The University of Maryland Strategic Partnership: MPowering the State, a formal collaboration between the University of Maryland, College Park, and the University of Maryland, Baltimore.

\section*{Declarations}

\begin{itemize}
\item \textbf{Competing interests:} The authors declare the following competing financial interest(s): P.T.\ is a consultant to Schrödinger, Inc.\ and is on their Scientific Advisory Board. P.T.\ is a co-founder of and holds equity in Emergente, Inc.
\item \textbf{Data availability:} The DUD-E actives and property-matched decoys used for benchmarking are publicly available at \url{https://dude.docking.org}. The receptor ensembles, docking results, and simulation data generated in this study are provided in the ANTIOCH-VS repository (\url{https://github.com/tiwarylab/antioch-vs}).
\item \textbf{Code availability:} ANTIOCH-VS is open-source and available at \url{https://github.com/tiwarylab/antioch-vs}.
\item Author contribution: H.H. and P.T. conceptualized the pipeline. H.H. implemented the structure-prediction, pocket-detection, and docking modules, generated the conformational ensembles, and carried out the docking benchmarks. V.A. developed the ligand-preparation and CTMD modules and the containerized HPC/Singularity execution, and performed the SPIB and CTMD analyses. H.H. wrote the original draft; V.A. and P.T. edited the manuscript. P.T. supervised the project. All authors reviewed and approved the final manuscript.
\end{itemize}

\begin{appendices}
\clearpage

\setcounter{figure}{0}
\renewcommand{\thefigure}{S\arabic{figure}}
\renewcommand{\theHfigure}{S\arabic{figure}}
\setcounter{table}{0}
\renewcommand{\thetable}{S\arabic{table}}
\renewcommand{\theHtable}{S\arabic{table}}

\section{Supplementary methods}\label{si:methods}

\subsection{JAK2 ensemble-generation parameters}

\textit{Structure prediction.} JAK2 kinase-domain conformers were generated with ColabFold~\cite{mirdita_colabfold_2022} using model type \texttt{alphafold2\_ptm} with no dropout, across eight reduced-MSA depths (\texttt{max\_msa} 8:16 through 1024:2048).

\textit{Feature selection and clustering.} Pairwise distances were computed with MDTraj~\cite{mcgibbon_mdtraj_2015} over the atoms defining the Dunbrack D1/D2 kinase-state coordinates---C$\alpha$ atoms of the $\alpha$C-helix/$\beta$3 region and the activation loop, together with sidechain atoms of the DFG motif (Asp157, Phe158, Gly159), the $\beta$3 lysine (Lys45), and the $\alpha$C glutamate (Glu61)---and ranked by coefficient of variation, keeping the top 200. The exact MDTraj atom selection (model numbering) was
\begin{quote}\footnotesize\ttfamily
((residue 41 to 69 or residue 129 to 187) and name CA) or (residue 157 and (name CB or name CG)) or (residue 158 and (name CZ or name CG)) or (residue 159 and name O) or (residue 61 and name CD) or (residue 45 and (name CB or name NZ))
\end{quote}
Regular-space clustering used a distance cutoff of $3.5$~\AA{}, batch size 100, and random seed 0.

\textit{Molecular dynamics.} Each cluster center was cleaned with PDBFixer and solvated in a truncated-octahedral TIP3P box with \texttt{tleap}, using the ff14SB (protein)~\cite{maier_ff14sb_2015}, GAFF2~\cite{wang_gaff_2004}, and TIP3P (water)~\cite{jorgensen_comparison_1983} force fields. Systems were energy-minimized (15{,}000 steps), heated from 0 to 100~K at 20~K\,ps$^{-1}$ and from 100 to 303~K at 2~K\,ps$^{-1}$, equilibrated for 1~ns under harmonic restraints on all backbone atoms and a further 1~ns under restraints on C$\alpha$ atoms only ($k = 5$~kcal\,mol$^{-1}$\,\AA$^{-2}$), and propagated for 50~ns of unbiased production per cluster center with the GPU \texttt{pmemd.cuda} engine of Amber~\cite{salomon-ferrer_amber_2013,case_ambertools_2023} patched with PLUMED~2.9.2~\cite{tribello_plumed_2014}. A 2~fs timestep was used, with temperature maintained at 303~K by a Langevin thermostat~\cite{loncharich_langevin_1992} (collision frequency 1~ps$^{-1}$) and pressure at 1~bar by an isotropic Berendsen barostat~\cite{berendsen_molecular_1984}, a 10~\AA{} nonbonded cutoff, and particle-mesh Ewald long-range electrostatics~\cite{darden_particle_1993}.

\textit{SPIB.} Trajectories from all cluster centers were pooled to train a two-dimensional SPIB model~\cite{wang_spib_2021,mehdi_accelerating_2022} over 20 time lags spaced log-uniformly from 0.5 to 10~ns; the final time lag, 8.77~ns, was chosen as the point of steepest drop in the number of predicted metastable states between consecutive lags (information-bottleneck regularization $\beta_{\mathrm{IB}} = 0.001$, learning rate $10^{-3}$). The resulting free-energy surface separates the DFG-in and DFG-out states.

\subsection{ADRB2 ensemble-generation parameters}
\textit{Structure prediction.} ADRB2 conformers were generated from sequence with ColabFold~\cite{mirdita_colabfold_2022} using model type \texttt{alphafold2\_ptm}, MSA mode \texttt{mmseqs2\_uniref\_env}, no dropout, ranked by pLDDT, sampling eight reduced-MSA depths (\texttt{max\_msa} 8:16, 16:32, 32:64, 64:128, 128:256, 256:512, 512:1024, and 1024:2048).

\textit{Feature selection.} All pairwise C$\alpha$--C$\alpha$ distances (MDTraj~\cite{mcgibbon_mdtraj_2015} selection \texttt{name CA}) were ranked by coefficient of variation and the top 200 retained.

\textit{Membrane molecular dynamics.} Each cluster center was truncated at the disordered third intracellular loop (ICL3, residues 200--231 in the model numbering) and embedded in a pre-oriented 1-palmitoyl-2-oleoyl-\emph{sn}-glycero-3-phosphocholine (POPC) bilayer with Packmol-Memgen~\cite{schott-verdugo_packmol-memgen_2019} in TIP3P water at 0.15~M KCl (20~\AA{} minimum water thickness, 15~\AA{} solute--boundary padding), giving $\sim$64{,}600-atom periodic systems ($83.7 \times 83.7 \times 111.2$~\AA{}). Topologies were built with \texttt{tleap} using the ff14SB (protein)~\cite{maier_ff14sb_2015}, Lipid21 (POPC)~\cite{dickson_lipid21_2022}, GAFF2~\cite{wang_gaff_2004}, and TIP3P (water)~\cite{jorgensen_comparison_1983} force fields with Joung--Cheatham ion parameters~\cite{joung_determination_2008}, the two native disulfides bonded explicitly. Systems were energy-minimized (two stages), heated to 303~K (two stages), and equilibrated under graduated backbone and then C$\alpha$ positional restraints, followed by 100~ns of unbiased production MD per center with the GPU \texttt{pmemd.cuda} engine of Amber~\cite{salomon-ferrer_amber_2013,case_ambertools_2023}: a 2~fs timestep with SHAKE-constrained bonds to hydrogen, a Langevin thermostat~\cite{loncharich_langevin_1992} at 303~K ($\gamma = 1$~ps$^{-1}$), a semi-isotropic constant-surface-tension barostat (1~bar, zero surface tension), a 10~\AA{} nonbonded cutoff with particle-mesh Ewald electrostatics, and coordinates saved every 10~ps.

\textit{Pocket-focused SPIB.} The State Predictive Information Bottleneck~\cite{wang_spib_2021,mehdi_accelerating_2022} was trained on the 136 pairwise C$\alpha$--C$\alpha$ distances among the 17 residues lining the 3NY8 orthosteric pocket (Trp78, Thr79, Asp82, Val83, Val86, Thr87, Phe162, Ser172, Ser173, Ser176, Trp255, Phe258, Phe259, Asn262, Tyr277, Asn281, Tyr285). SPIB used a time lag of 1{,}708 frames (10~ps/frame; 17.1~ns), a two-dimensional reaction coordinate, encoder hidden layers of 32 and 128 neurons, an information-bottleneck regularization $\beta_{\mathrm{IB}}=0.05$, and Adam optimization (learning rate $10^{-4}$, batch size 512, up to 8 label-refinement rounds, random seed 42), resolving the binding-site conformations into seven metastable states.

\section{Benchmark interpretation across docking engines}\label{si:benchmark}

The absolute DUD-E enrichment we report should be read with care, because our setting differs from the original DUD-E study~\cite{mysinger_dude_2012} in two ways that both bear on enrichment. First, the docking programs are not the same. DUD-E was constructed and validated with DOCK3.6, of the UCSF DOCK3 lineage, whose physics-based scoring (solvent-excluded-volume ligand desolvation with a precomputed electrostatic potential) is distinct from the grid-based force-field Grid score of DOCK6~\cite{balius_dock_2024} used here; despite the shared name these are different programs with different scoring functions, and even head-to-head on identical decoy sets their enrichment diverges (DOCK3.7 enriches better than DOCK6 on DUDE-Z and SB2012, while DOCK6 surpasses it once post-docking torsion minimization is added~\cite{balius_dock_2024}). Second, enrichment against property-matched decoys has well-documented liabilities: DUD-E decoys are matched on a few physical properties while being topologically dissimilar to the actives, which can leave them separable by artifacts of the selection procedure rather than by genuine binding complementarity~\cite{chen_hidden_2019}; such sets neither span the full chemical spectrum of a prospective library nor protect against false confidence when enrichment is optimized against them~\cite{stein_property-unmatched_2021}; and for GPCRs in particular the decoys are matched only in low-dimensional properties, so apparent enrichment is highly sensitive to how they are generated~\cite{weiss_gpcr-bench_2016}. Enrichment on these sets also carries substantial statistical variability, and differences that appear large can shrink under resampling~\cite{stein_property-unmatched_2021}.

\begin{figure}[htbp]
    \centering
    \includegraphics[width=\textwidth]{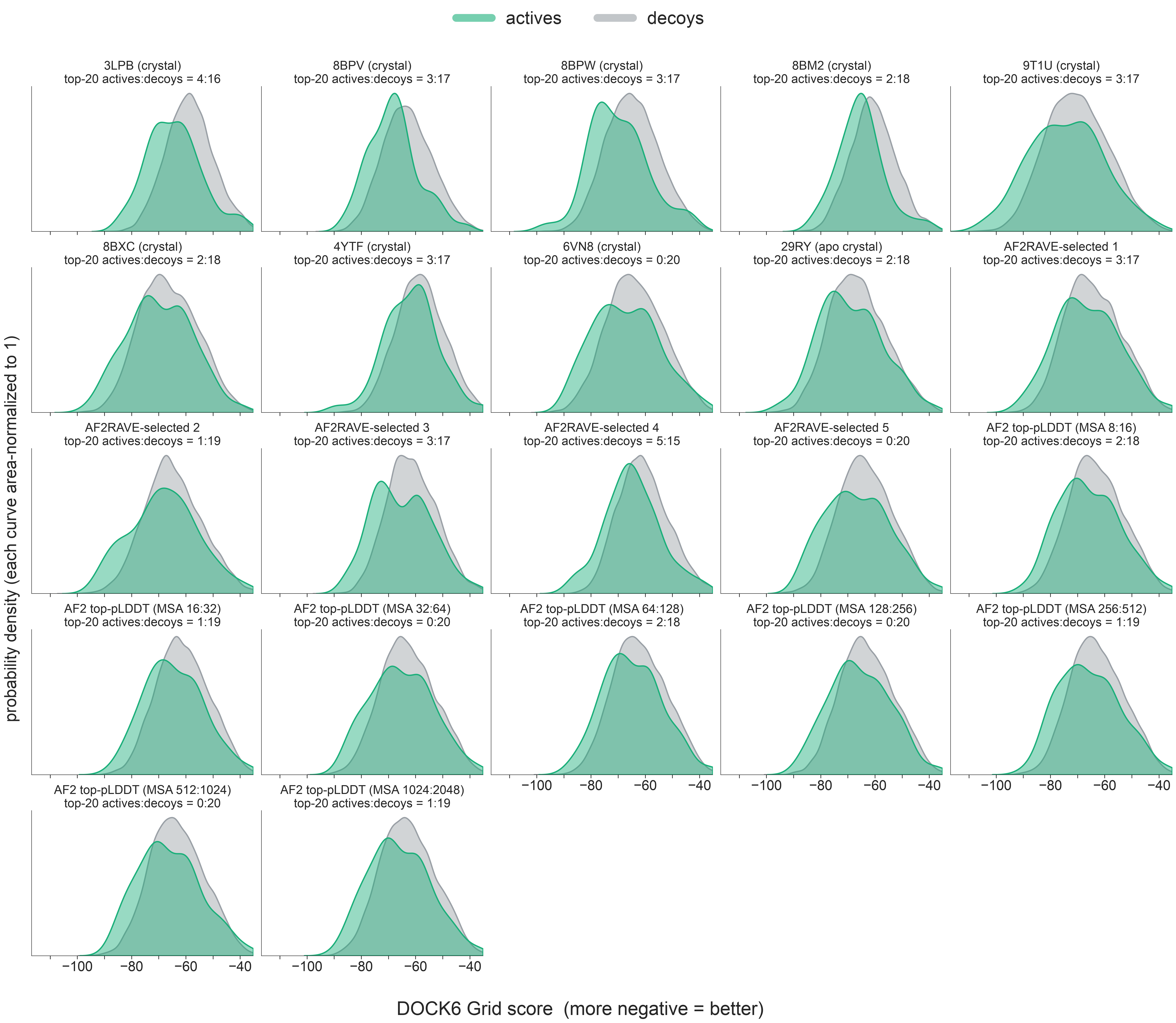}
    \caption{DOCK6 Grid-score distributions of DUD-E actives (green) and decoys (grey) for every JAK2 receptor: the eight inhibitor-bound crystals, the apo crystal (29RY), the five AF2RAVE-selected structures, and the eight AF2 top-pLDDT models (one per reduced-MSA depth, labelled by its \texttt{max\_seq:max\_extra\_seq} setting). Each panel title gives that receptor's top-20 active:decoy split. In contrast to ADRB2 (Fig.~\ref{si:adrb2_dockscore}), the JAK2 actives and decoys overlap heavily and no structure---crystal, AF2RAVE, or AF2---separates the two classes by a large Grid-score margin; the activation-loop/DFG inhibitor reference 3LPB shows the clearest active-enriched shoulder (top-20 4:16), and the top of the ranking is never dominated by actives. JAK2 is thus a genuinely harder benchmark whose decoys are docked competitively with the actives, underscoring that the uniformly high ADRB2 top-20 rates reflect the ADRB2 DUD-E decoy set rather than a general property of the pipeline. \emph{Note:} each curve is a kernel-density estimate area-normalized to~1 independently, so the vertical axis is a probability density (arbitrary units) whose heights do \emph{not} reflect the $\sim$59:1 decoy-to-active abundance ($\sim$107 actives vs.\ $\sim$6{,}330 decoys); the comparison is of distribution shape and location, with abundance carried by the per-panel top-20 ratio. The score axis spans the full favourable (negative) range and is trimmed only at the extreme positive (failed/unbound-pose) tail.}
    \label{si:jak2_dockscore}
\end{figure}

\begin{figure}[htbp]
    \centering
    \includegraphics[width=\textwidth]{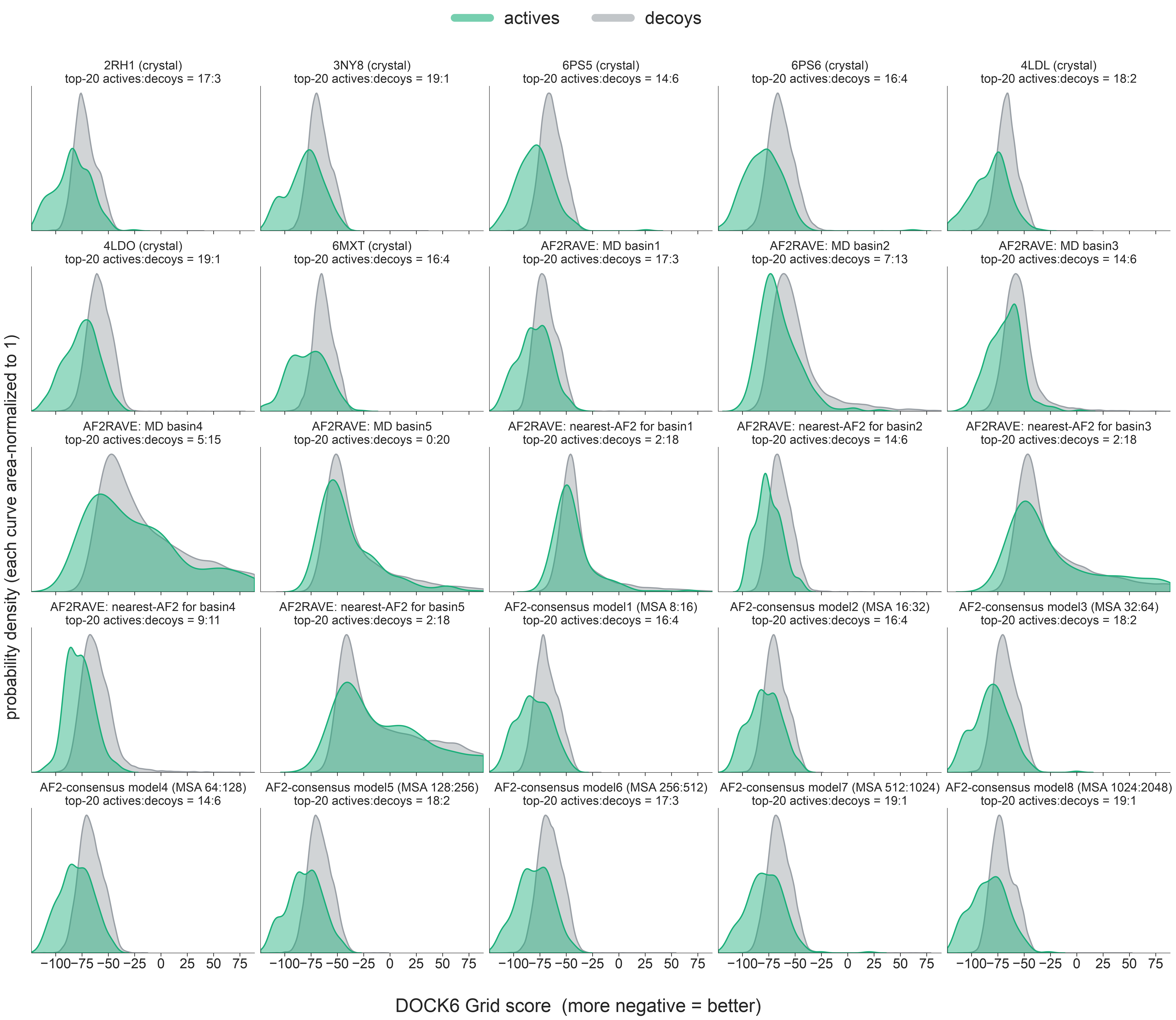}
    \caption{DOCK6 Grid-score distributions of DUD-E actives (green) and decoys (grey) for every ADRB2 receptor, ordered as the seven ligand-bound/apo crystals, the five AF2RAVE pocket-SPIB basins represented by their MD-minimum snapshot and by their geometrically nearest AF2 prediction, and the eight AF2-consensus top-pLDDT models (one per reduced-MSA depth, labelled by its \texttt{max\_msa} setting). Each panel title gives that receptor's top-20 active:decoy split (the number of true actives among its 20 best-scoring compounds). Wherever the pocket is well formed---every crystal, every AF2-consensus model, and the AF2RAVE basin-1/2/3 conformations---the actives are shifted to systematically more favourable (more negative) Grid scores than the decoys, by a median of roughly 11--16~kcal/mol, so the top of the ranking fills with actives regardless of whether the structure came from crystallography, standard AF2, or AF2RAVE. Separation collapses only for the distorted or coverage-limited pockets (the basin-4 and basin-5 MD minima and the nearest-AF2 predictions for basins~1, 3, and 5), which also fail the docking-rate filter (Table~\ref{si:adrb2_rates}). This systematic, structure-source-independent separation indicates that the ADRB2 DUD-E decoys are readily down-ranked by DOCK6 once a physically reasonable orthosteric pocket is present, and accounts for the uniformly high top-20 hit rates on this target. \emph{Note:} each curve is a kernel-density estimate area-normalized to~1 independently, so the vertical axis is a probability density (arbitrary units) and the curve heights do \emph{not} reflect the $\sim$64:1 decoy-to-active abundance ($\sim$231 actives vs.\ $\sim$14{,}700 decoys); the figure compares the \emph{shape and location} of the two score distributions, not their counts, with abundance information carried instead by the per-panel top-20 ratio.}
    \label{si:adrb2_dockscore}
\end{figure}

These effects are visible directly in our data. Comparing the same receptor structures and decoy sets across the two engines (Table~\ref{si:engine}), the difficulty ordering of our two targets \emph{inverts}: DOCK3.6 enriched JAK2 strongly and ADRB2 only modestly, whereas DOCK6 does the opposite. The per-receptor Grid-score distributions (SI Figs.~\ref{si:jak2_dockscore},~\ref{si:adrb2_dockscore}) show why---under DOCK6 the ADRB2 actives are separated from the property-matched decoys at every well-formed pocket (crystal, AF2-consensus, or AF2RAVE), with a median Grid-score margin of $\approx$11--16~kcal/mol, so the top of nearly every ranking fills with actives (e.g.\ a plain AF2 consensus for which all twenty top-ranked ADRB2 compounds are actives), while the JAK2 actives and decoys overlap heavily for every receptor. The most parsimonious reading is that the ADRB2 DUD-E decoys are easy for DOCK6 to down-rank once a physically reasonable orthosteric pocket is present, not that any single receptor is uniquely performant. Because the absolute magnitude of DUD-E enrichment reflects the scoring-function/decoy-set pairing more than an intrinsic property of the target or of the generated ensemble, we base the conclusions in the main text on within-target, same-decoy comparisons (AF2RAVE-selected ensembles against the structure-free baselines and the ligand-bound crystals), where these decoy-set idiosyncrasies largely cancel, and on physics-based CTMD triage. The goal of this work is to demonstrate that the pipeline runs end to end, not to benchmark docking-software accuracy or DUD-E itself.

\begin{table}[htbp]
    \centering
    \caption{The same JAK2 and ADRB2 receptor structures and DUD-E decoy sets give opposite difficulty orderings under the two docking engines. DUD-E/DOCK3.6 values are from the original study's per-target table~\cite{mysinger_dude_2012} (DUD-E overall mean over 102 targets: adjusted LogAUC 24.4, EF1 19.8, ROC AUC 76.5). DOCK6 values are from this work, for the identical crystal structure used by DUD-E (JAK2 3LPB, ADRB2 3NY8) and for the AF2RAVE-selected consensus. All three metrics---DUD-E adjusted LogAUC, the ROC-based enrichment factor at 1\% (EF1), and ROC AUC---share definitions across the two studies and are compared directly.}
    \label{si:engine}
    \begin{tabular}{llccc}
    \toprule
    Target & Receptor / engine & Adj.\ LogAUC (\%) & EF1 & ROC AUC (\%) \\
    \midrule
    JAK2  & 3LPB crystal --- DOCK3.6 (DUD-E)      & 28.6 & 23.4 & 81.7 \\
    JAK2  & 3LPB crystal --- DOCK6 (this work)    & 13.0 &  9.3 & 67.8 \\
    JAK2  & AF2RAVE ($n=5$) --- DOCK6 (this work) &  8.2 &  7.5 & 58.9 \\
    \addlinespace
    ADRB2 & 3NY8 crystal --- DOCK3.6 (DUD-E)      & 11.7 &  3.9 & 69.3 \\
    ADRB2 & 3NY8 crystal --- DOCK6 (this work)    & 30.1 & 30.6 & 75.2 \\
    ADRB2 & AF2RAVE ($n=5$) --- DOCK6 (this work) & 32.1 & 33.3 & 78.3 \\
    \bottomrule
    \end{tabular}

    \begin{flushleft}\footnotesize
    EF1 is the ROC-based enrichment factor at 1\% decoy recall---equivalently, the percent of actives recovered once 1\% of the decoys have been recovered---which our pipeline computes identically to DUD-E~\cite{mysinger_dude_2012}, so the values are directly comparable. The flip is seen in all three metrics: DOCK3.6 enriches JAK2 far more strongly than ADRB2, while DOCK6 does the reverse.
    \end{flushleft}
\end{table}

\clearpage

\section{Supplementary figures and tables}\label{secSI}

\begin{figure}[htbp]
    \centering
    \includegraphics[width=0.7\textwidth]{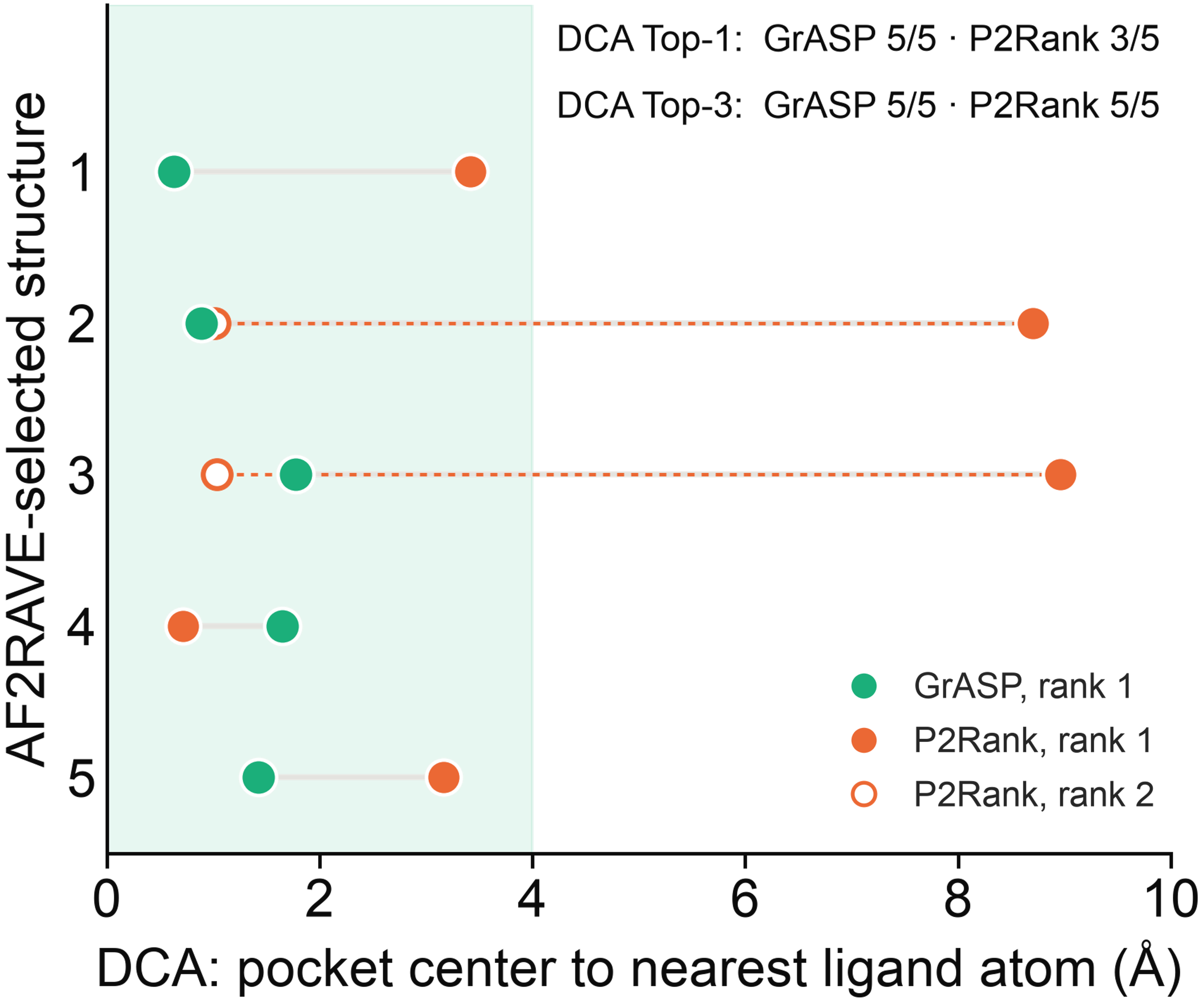}
    \caption{Pocket-detection accuracy on the five AF2RAVE-selected JAK2 structures (GrASP vs.\ P2Rank, scored by distance to the nearest crystal-ligand atom).}
    \label{si:pocket}
\end{figure}

\begin{figure}[htbp]
    \centering
    \includegraphics[width=0.49\textwidth]{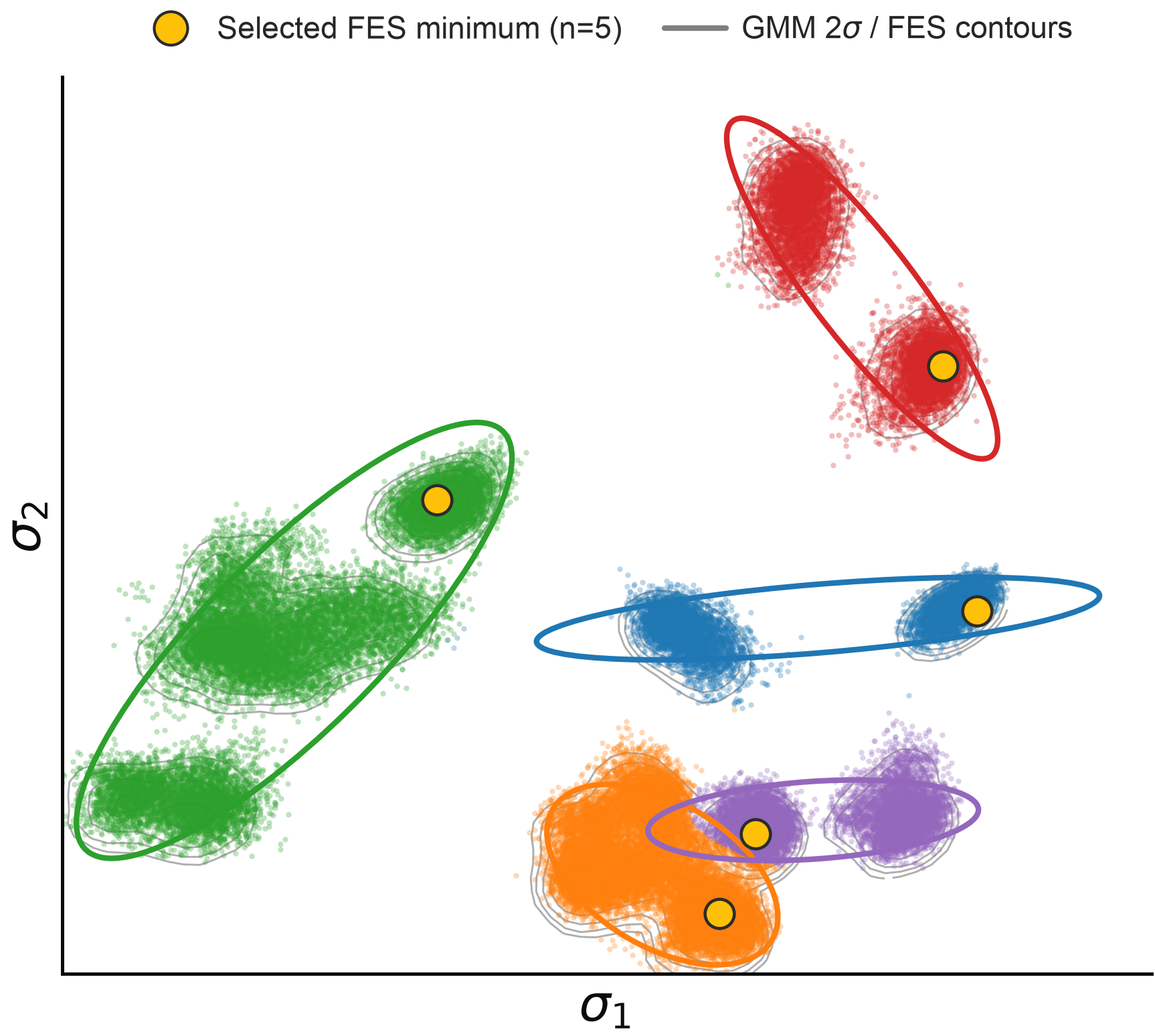}\hfill
    \includegraphics[width=0.49\textwidth]{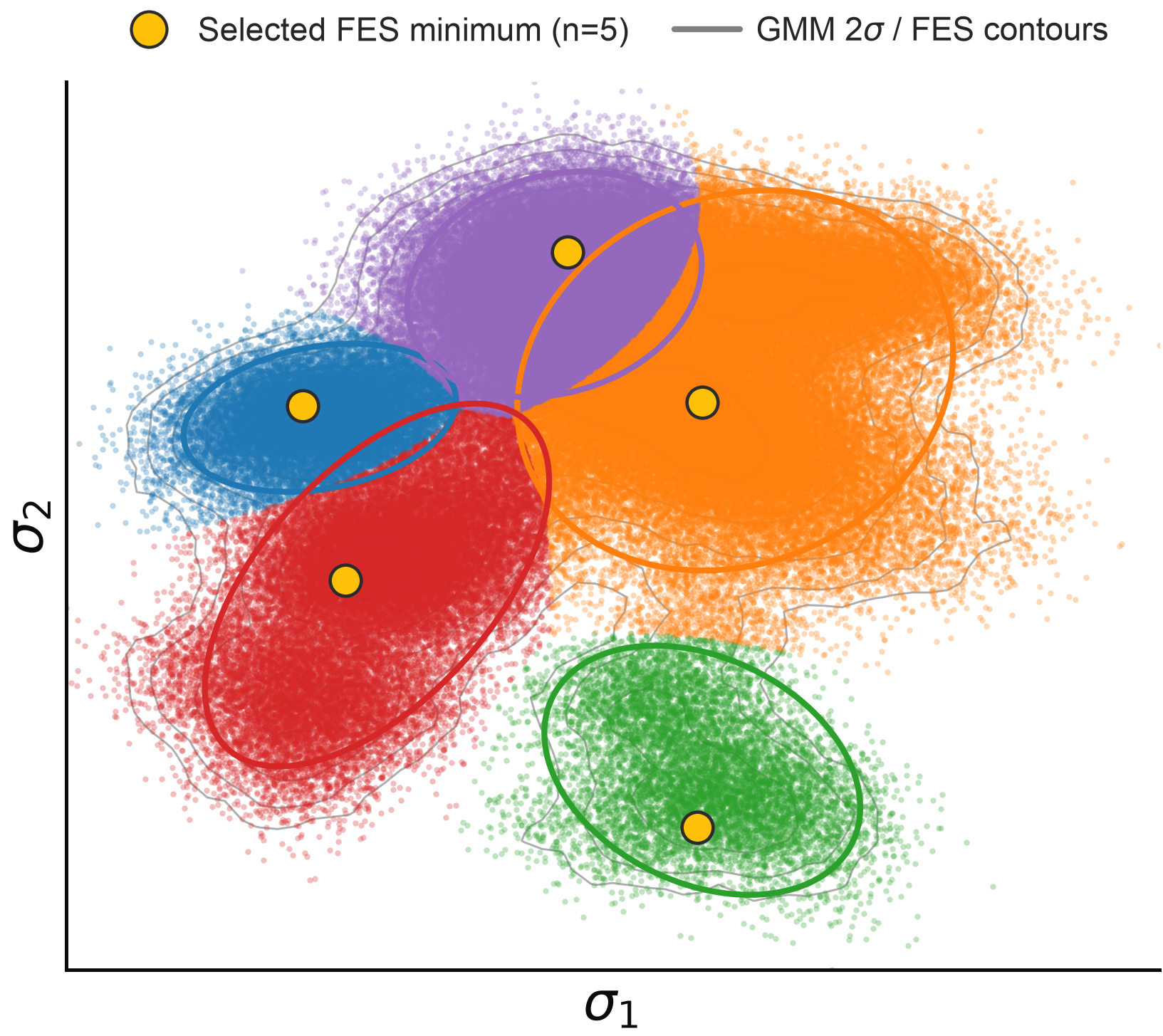}
    \caption{Selecting the docking conformations from the SPIB free-energy surface with a Gaussian mixture model (GMM), for (left) JAK2 and (right) ADRB2. Points are the unbiased-MD frames in the two-dimensional SPIB space ($\sigma_1$, $\sigma_2$), coloured by their GMM component; each component's $2\sigma$ ellipse is drawn in the matching colour and the grey lines are free-energy contours. A five-component GMM (\texttt{scikit-learn} \texttt{GaussianMixture}, \texttt{random\_state}=9) was fit to each surface, and the lowest-free-energy frame of each component (gold circle) was taken as a representative free-energy minimum, giving the five docking conformations per target. For JAK2 these five minima map to the five AF2RAVE-selected receptors (the $n=5$ consensus); for ADRB2 they are the five MD minima (md1--md5) that, with their nearest AF2 predictions, enter the docking-rate screen (Table~\ref{si:adrb2_rates}). The JAK2 surface shows well-separated basins, whereas the ADRB2 pocket-SPIB surface is more collapsed, with adjacent, overlapping components---consistent with the narrower conformational spread of the orthosteric pocket.}
    \label{si:spib_gmm}
\end{figure}

\begin{figure}[htbp]
    \centering
    \includegraphics[width=\textwidth]{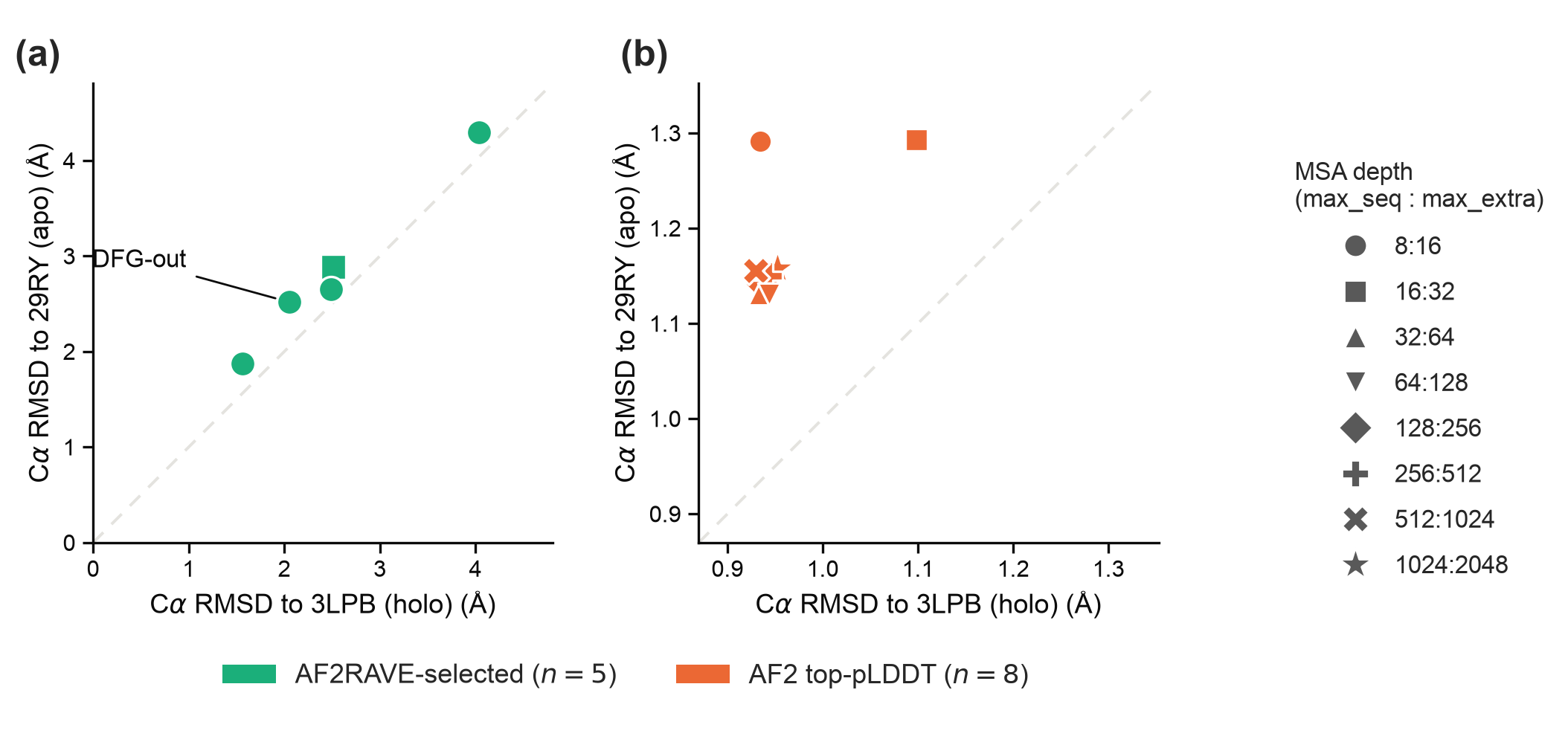}
    \caption{C$\alpha$ RMSD of each sequence-derived JAK2 model to the two reference crystals---3LPB (inhibitor-bound, $x$) and 29RY (apo, $y$)---the quantitative companion to the structural overlays in Fig.~\ref{fig:enrichment}. (a) The five AF2RAVE-selected conformations (green); (b) the eight AF2 top-pLDDT models (orange), on a zoomed axis (note the different scale). Marker shape encodes the reduced-MSA depth the model came from (right legend, \texttt{max\_seq}:\texttt{max\_extra\_seq}); the dashed diagonal marks equal distance to the two crystals. The AF2RAVE ensemble spreads 1.6--4.0~\AA{} from 3LPB---including the DFG-out member (labelled)---sampling conformations no single crystal represents, whereas the AF2 top-pLDDT models collapse into a tight cluster 0.9--1.1~\AA{} from 3LPB regardless of MSA depth. RMSDs are full-structure values over all matched C$\alpha$ (PyMOL \texttt{align}, \texttt{cycles}=0, no outlier trimming), computed on the relaxed structures.}
    \label{si:ref_rmsd}
\end{figure}

\begin{table}[htbp]
    \centering
    \caption{Per-receptor retrospective DUD-E enrichment for the JAK2 kinase domain (107 actives, 6{,}500 property-matched decoys; DOCK6), corresponding to Fig.~\ref{fig:enrichment}C,D. Rows follow the Fig.~\ref{fig:enrichment} ordering: the crystals and the AF2RAVE consensus ranked by adjusted LogAUC, with the single- and consensus-AF2 baselines pinned below. Values are adjusted LogAUC and EF1 (\%).}
    \label{si:jak2_enrichment}
    \begin{tabular}{llcc}
    \toprule
    Receptor & Source & Adj.\ LogAUC & EF1 \\
    \midrule
    3LPB                        & inhibitor-bound crystal (DUD-E ref.) & 12.99 & 9.35 \\
    8BPV                        & inhibitor-bound crystal              & 9.66  & 4.72 \\
    8BPW                        & inhibitor-bound crystal              & 9.37  & 3.74 \\
    8BM2                        & inhibitor-bound crystal              & 8.70  & 5.61 \\
    AF2RAVE consensus ($n=5$)   & sequence-derived (AF2RAVE)           & 8.25  & 7.48 \\
    9T1U                        & inhibitor-bound crystal              & 7.60  & 4.67 \\
    8BXC                        & inhibitor-bound crystal              & 6.50  & 5.61 \\
    4YTF                        & inhibitor-bound crystal              & 6.47  & 3.74 \\
    6VN8                        & inhibitor-bound crystal              & 6.44  & 3.74 \\
    29RY                        & apo crystal                          & 5.25  & 2.80 \\
    \addlinespace
    AF2, single model ($n=1$)   & sequence-derived (AF2)               & 6.28  & 3.74 \\
    AF2 top-pLDDT consensus ($n=8$) & sequence-derived (AF2)           & 5.98  & 3.74 \\
    \bottomrule
    \end{tabular}

    \begin{flushleft}\footnotesize
    Enrichment is computed on the compounds each receptor successfully scored (Methods). The seven inhibitor-bound crystals other than 3LPB span adjusted LogAUC 6.44--9.66 and EF1 3.74--5.61 (the ``inhibitor-bound-crystal band'' in Fig.~\ref{fig:nocrystal}A). The AF2RAVE and AF2 rows are RRF consensuses over their member structures; paired-bootstrap confidence intervals and significance for the no-inhibitor-crystal comparison are given in Table~\ref{si:stats}.
    \end{flushleft}
\end{table}

\begin{table}[htbp]
    \centering
    \caption{Paired-bootstrap enrichment statistics for the JAK2 no-inhibitor-crystal scenarios of Fig.~\ref{fig:nocrystal}A (JAK2 DUD-E benchmark, DOCK6). Values are adjusted LogAUC and EF1; the AF2RAVE consensus ($n = 5$) is the reference against which each source is compared.}
    \label{si:stats}
    \begin{tabular}{lccc}
    \toprule
    Receptor source & Value [95\% CI] & $\Delta$ vs.\ AF2RAVE$^{a}$ [95\% CI] & $p^{b}$ \\
    \midrule
    \multicolumn{4}{l}{\textbf{Adjusted LogAUC}} \\
    \quad Apo crystal (29RY)             & 5.24 [1.71, 8.97]   & $+3.00$ [1.32, 4.80]          & $<0.001$ \\
    \quad AF2, single model ($n = 1$)    & 6.28 [2.76, 10.16]  & $+1.96$ [0.71, 3.43]          & $0.001$  \\
    \quad AF2 consensus ($n = 8$)        & 5.96 [2.41, 9.73]   & $+2.27$ [1.22, 3.45]          & $<0.001$ \\
    \quad AF2RAVE consensus ($n = 5$)    & 8.23 [4.16, 12.80]  & --- (ref.)                    & ---      \\
    \quad Inhibitor-bound crystal (3LPB) & 12.99 [8.76, 17.71] & $-4.75$ [$-7.41$, $-2.18$]    & $1.000$  \\
    \addlinespace
    \multicolumn{4}{l}{\textbf{EF1}} \\
    \quad Apo crystal (29RY)             & 2.80 [0.00, 6.54]   & $+4.67$ [0.93, 9.10]          & $0.017$  \\
    \quad AF2, single model ($n = 1$)    & 3.74 [0.00, 7.48]   & $+3.74$ [0.00, 7.48]          & $0.045$  \\
    \quad AF2 consensus ($n = 8$)        & 3.74 [0.93, 7.48]   & $+3.74$ [0.00, 7.48]          & $0.042$  \\
    \quad AF2RAVE consensus ($n = 5$)    & 7.48 [2.04, 12.15]  & --- (ref.)                    & ---      \\
    \quad Inhibitor-bound crystal (3LPB) & 9.35 [3.74, 14.95]  & $-1.87$ [$-7.48$, $2.80$]     & $0.818$  \\
    \bottomrule
    \end{tabular}

    \begin{flushleft}\footnotesize
    $^{a}$~$\Delta$ is the AF2RAVE consensus value minus the listed source (positive favors AF2RAVE), with its 95\% CI from the paired bootstrap over the common ligand split.\\
    $^{b}$~One-sided paired-bootstrap probability that the AF2RAVE consensus does not exceed the source; ``---'' for the reference row.\\
    Point estimates and intervals are computed on the ligands scored by every DOCK6 receptor (the shared set: 107 actives and 6,327 of the 6,500 DUD-E decoys), so values differ by $<0.1$ from the per-receptor estimates in Fig.~\ref{fig:enrichment}.\\
    \textit{Inhibitor-bound-crystal band} (the seven inhibitor-bound crystals other than 3LPB): adjusted LogAUC 6.44--9.66, EF1 3.74--5.61.
    \textit{Per-model comparison} (Fig.~\ref{fig:nocrystal}B,C), AF2RAVE-selected ($n = 5$) vs.\ AF2 top-pLDDT ($n = 8$), two-sided Mann--Whitney: LogAUC $p = 0.093$, EF1 $p = 0.003$.
    \textit{Cluster-center subset sweep} (Fig.~\ref{fig:nocrystal}D,E; all 3,003 five-of-15 consensuses): the SPIB-selected set ranks in the 100th percentile (LogAUC) and 99th percentile (EF1); the all-15 consensus gives LogAUC 6.75, EF1 4.67.
    \end{flushleft}
\end{table}

\begin{figure}[htbp]
    \centering
    \includegraphics[width=\textwidth]{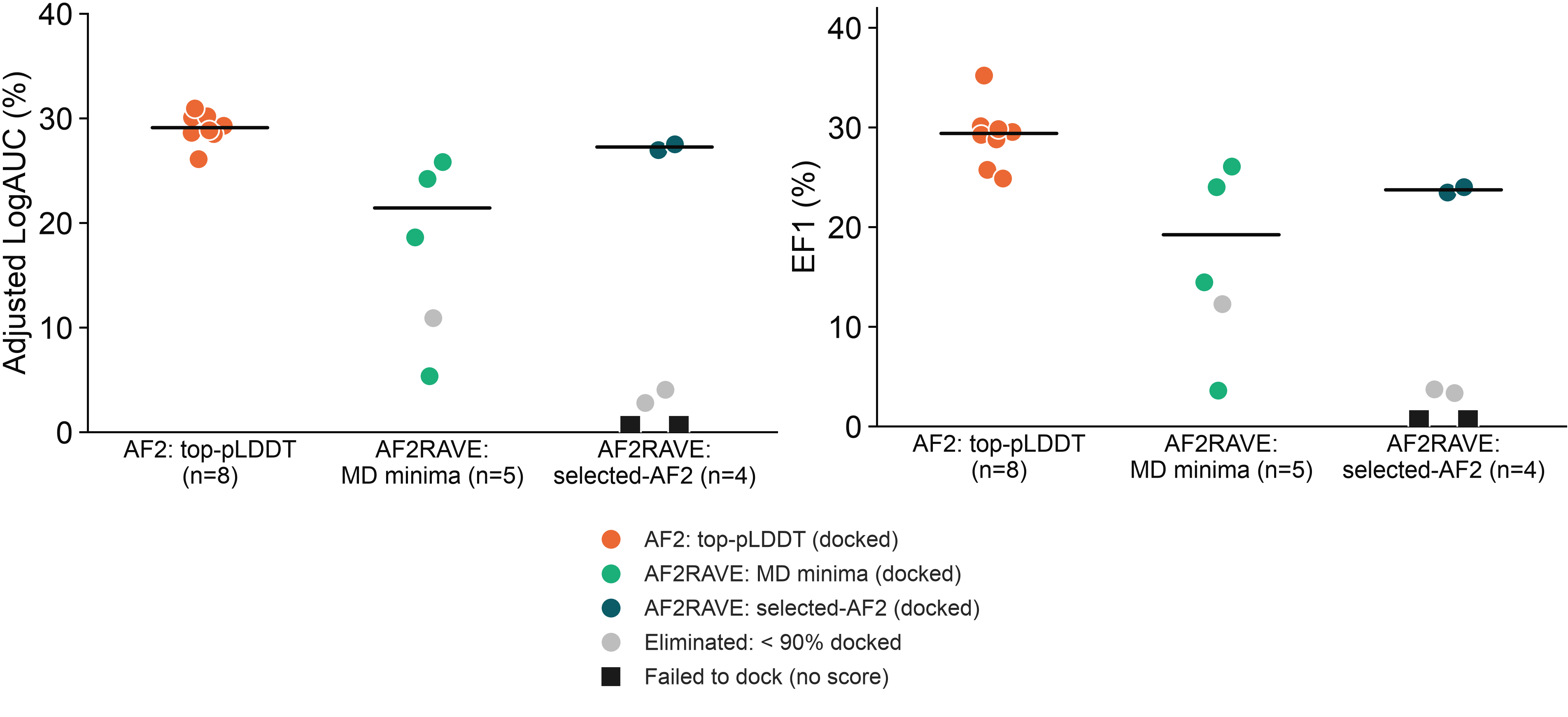}
    \caption{Per-conformation retrospective DOCK6 enrichment for the ADRB2 AF2RAVE candidates (adjusted LogAUC, left; EF1, right). Each point is one receptor conformation: the eight AF2 top-pLDDT models (orange), the five pocket-SPIB MD-minimum snapshots (green), and the geometrically nearest AF2 predictions selected alongside them (teal). Colour marks the docking outcome used for selection: group colour, kept in the AF2RAVE consensus (docked $\geq 90\%$ of the library); grey, eliminated for docking $<90\%$ (the basin-4 MD minimum and the AF2 prediction nearest basin 3); black at the axis floor, failed DOCK6 sphere generation with no score (the AF2 predictions nearest basins 4 and 5). Bars, group medians over the kept members. The basin-5 MD minimum is retained by the docking-rate filter but is the weakest kept conformation (adjusted LogAUC 5.4, EF1 3.6), motivating the $n=5$ (excluding it) and $n=6$ (including it) consensus rows in Fig.~\ref{fig:adrb2}E.}
    \label{si:adrb2_selection}
\end{figure}

\begin{table}[htbp]
    \centering
    \caption{Docking-rate and retrospective-enrichment accounting for the ten ADRB2 AF2RAVE docking candidates (five pocket-SPIB basins $\times$ \{MD minimum, nearest AF2\}; DOCK6). Docking rate is the fraction of the submitted library that produced a valid pose on that conformation; conformations below the 90\% threshold are excluded from the AF2RAVE consensus. Enrichment is single-receptor retrospective adjusted LogAUC and EF1.}
    \label{si:adrb2_rates}
    \begin{tabular}{llccccl}
    \toprule
    Basin & Structure & Docking rate (\%) & Adj.\ LogAUC & EF1 & Decision \\
    \midrule
    1 & MD minimum  & 97.9 & 25.8 & 26.1 & kept \\
    1 & nearest AF2 & 97.7 & 27.5 & 24.0 & kept \\
    2 & MD minimum  & 96.4 & 18.6 & 14.5 & kept \\
    2 & nearest AF2 & 97.9 & 27.0 & 23.5 & kept \\
    3 & MD minimum  & 97.5 & 24.2 & 24.0 & kept \\
    3 & nearest AF2 & 77.2 & 4.1  & 3.4  & dropped ($<90\%$) \\
    4 & MD minimum  & 68.1 & 10.9 & 12.3 & dropped ($<90\%$) \\
    4 & nearest AF2 & ---  & ---  & ---  & failed (sphere-gen) \\
    5 & MD minimum  & 90.3 & 5.4  & 3.6  & kept \\
    5 & nearest AF2 & ---  & ---  & ---  & failed (sphere-gen) \\
    \bottomrule
    \end{tabular}

    \begin{flushleft}\footnotesize
    Submitted library: the 231 active and 14{,}993 decoy unique DUD-E ligands expanded to 447 active and 15{,}255 decoy protomer/tautomer states (15{,}702 states); docking rate $=$ scored states $/$ submitted states, computed per conformation from its DOCK6 output.
    Basins are numbered by their MD minimum (basin $N$ = the SPIB basin whose MD free-energy minimum is minimum $N$); ``nearest AF2'' is the geometrically nearest AF2 prediction to that basin minimum.
    For basins 4 and 5 the nearest AF2 predictions failed sphere generation; fallback predictions from the same basins were tried but not used (basin-4 fallback docked only 62.0\%, below threshold; the basin-5 fallback fell in a different basin and was discarded), so neither basin contributes an AF2 proxy.
    The kept conformations form the AF2RAVE-selected consensus; excluding the basin-5 MD minimum gives the $n=5$ consensus and including it the $n=6$ consensus (Fig.~\ref{fig:adrb2}E).
    \end{flushleft}
\end{table}
\newpage
\begin{figure}[htbp]
    \centering
    \includegraphics[width=0.9\textwidth]{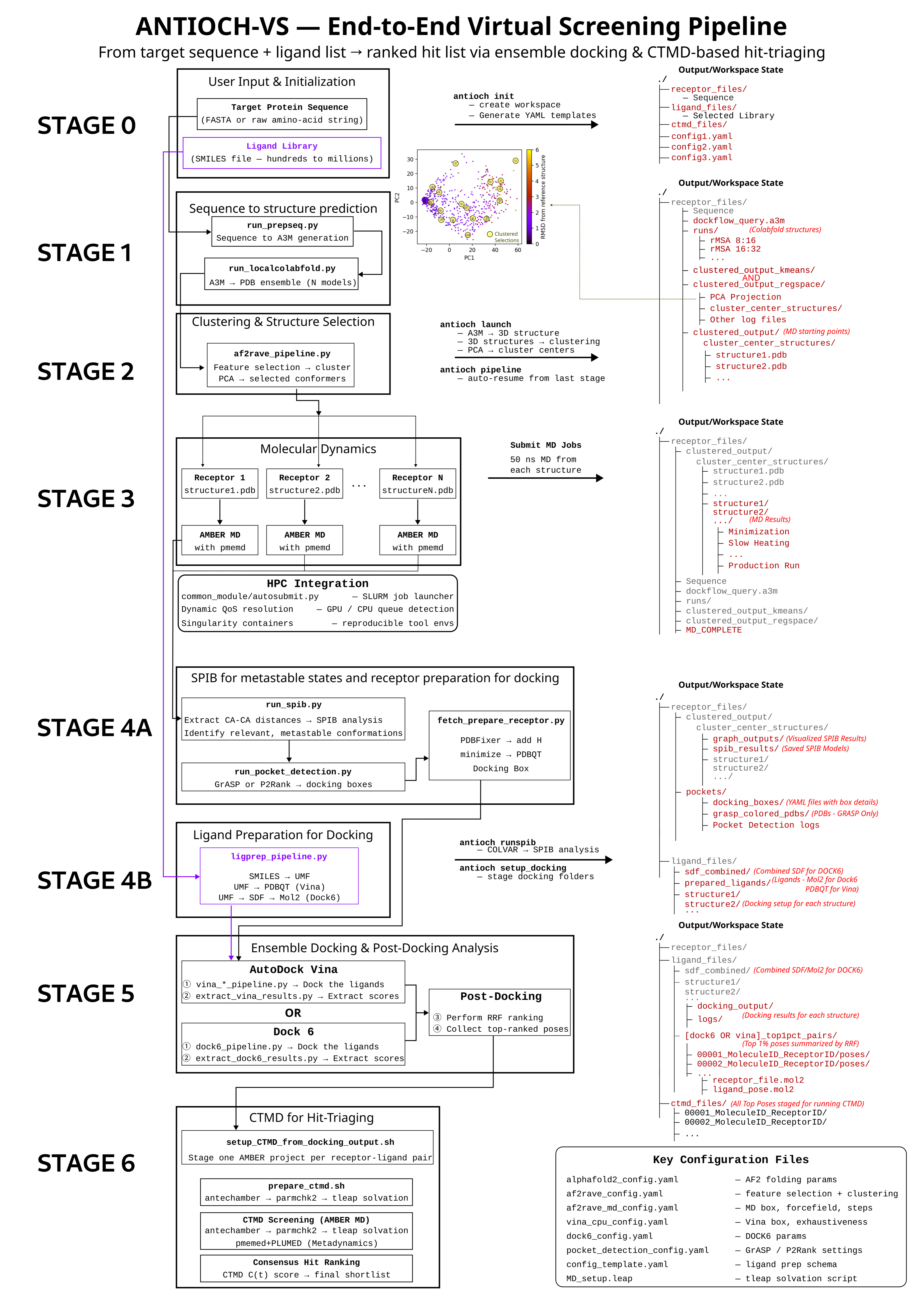}
    \caption{The complete pipeline from start to finish. Stage 0 is the only point where user input is required. The left panel describes each program (available on our GitHub repository) and its role in the pipeline. The right panel shows a running description of the output (folders and files created). The red coloured files are the files created in that step. Each step can be run independently using a specific command. ``\textbf{antioch pipeline}'' is a placeholder that continues the pipeline from any point after initialization.}
    \label{si:the_full_pipeline}
\end{figure}

\begin{figure}[htbp]
    \centering
    \includegraphics[width=\textwidth]{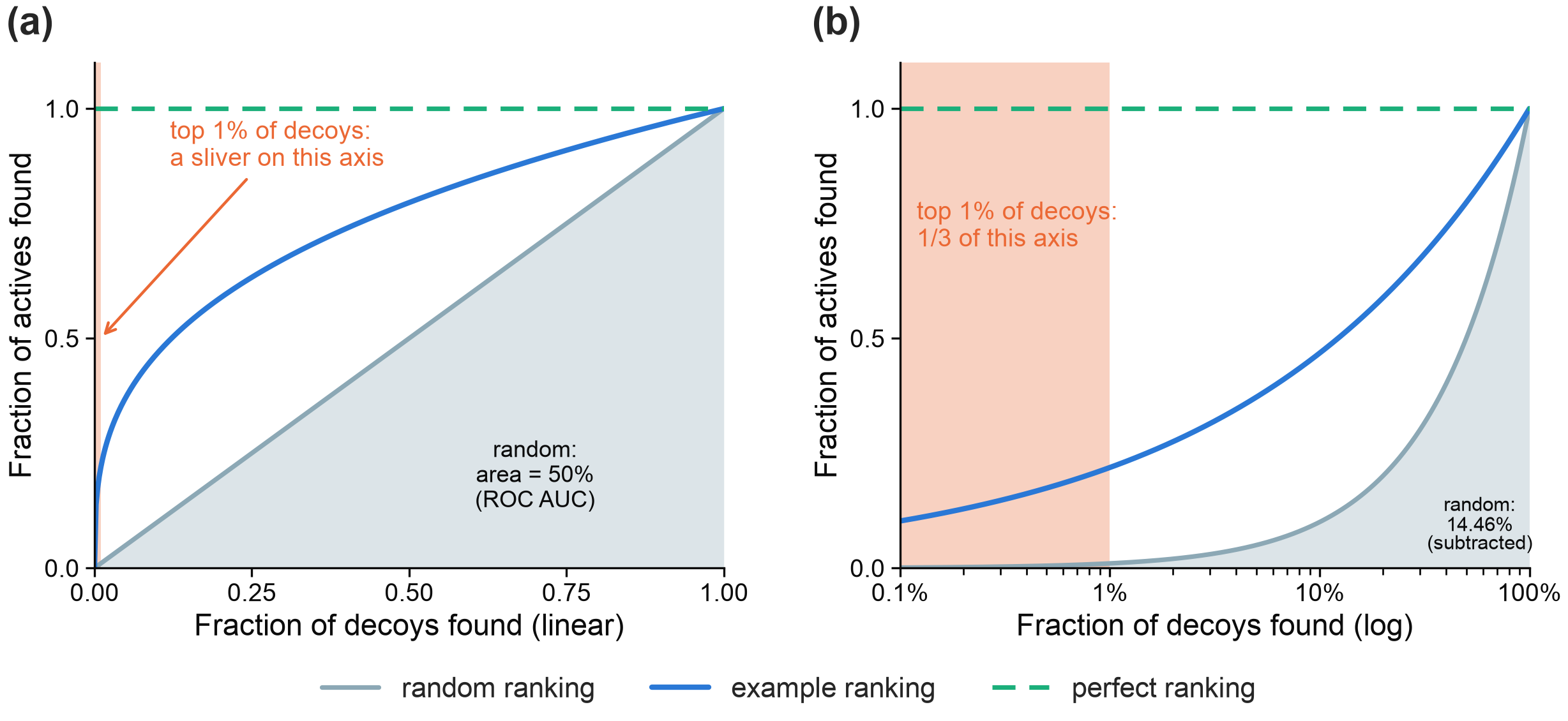}
    \caption{How the adjusted LogAUC reported throughout this work is defined, following the DUD-E convention~\cite{mysinger_dude_2012} as implemented in the pipeline's enrichment code. Both panels show the same two illustrative rankings (a better-than-random example and the random and perfect references). \textbf{(a)} On a linear decoy-recall axis the random ranking (diagonal) encloses $50\%$ of the area---the ordinary ROC AUC---and the top-1\% early-enrichment window where EF$_{1\%}$ is read is a thin sliver at the left edge, so this axis gives early and late enrichment equal visual weight. \textbf{(b)} On a logarithmic decoy-recall axis (three decades, $0.1\%$ to $100\%$) the top-1\% window instead occupies a third of the axis, emphasizing early enrichment; the random ranking now encloses $14.46\%$ of the area. Adjusted LogAUC is this semilog area, normalized so a perfect ranking scores $100\%$, minus the $14.46\%$ random baseline, so that random ranking scores $0$. The same $0.1\%$ minimum decoy fraction and $1\%$ EF cutoff are used for every receptor.}
    \label{si:logauc_explainer}
\end{figure}

\begin{figure}[htbp]
    \centering
    \includegraphics[width=\textwidth]{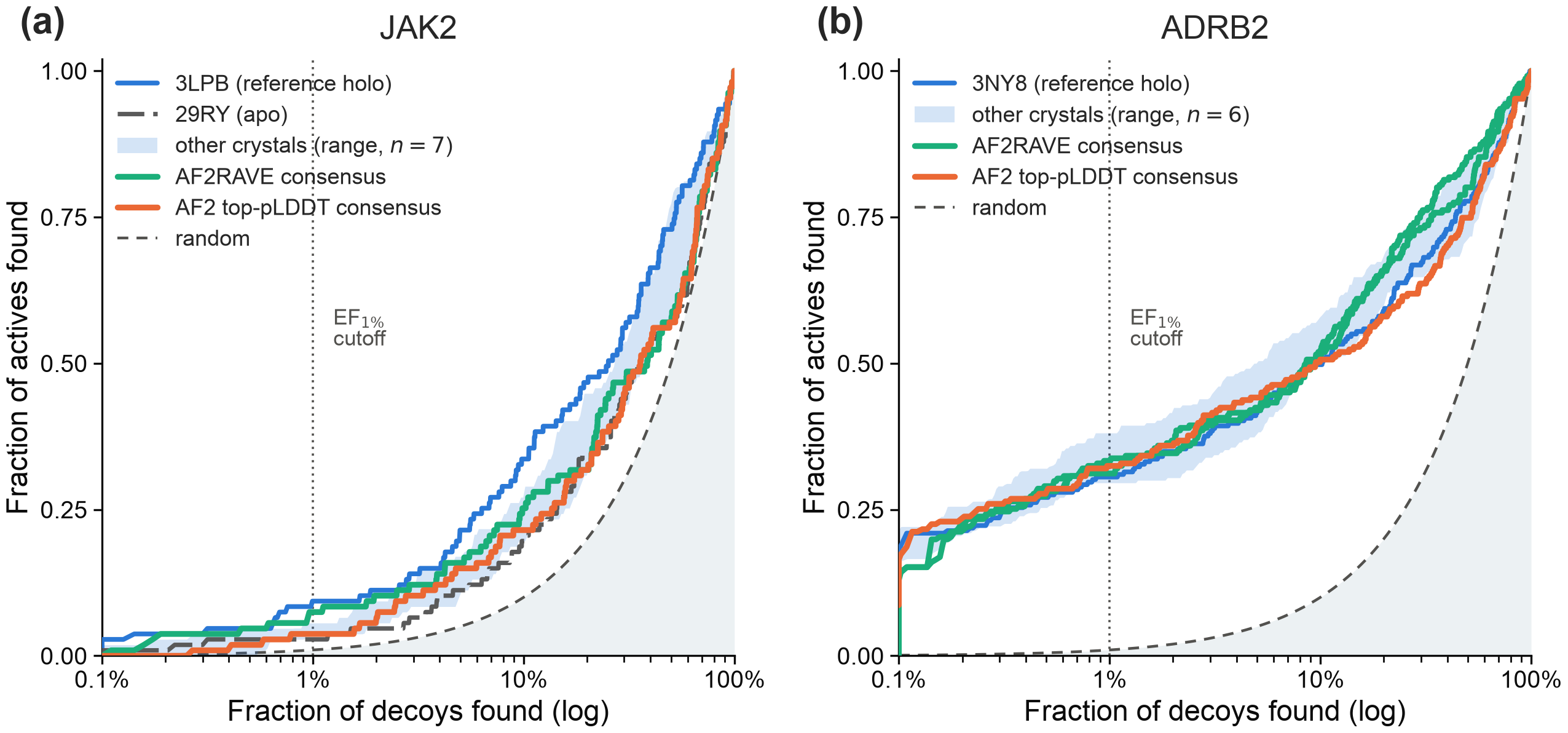}
    \caption{ROC curves (fraction of DUD-E actives recovered versus fraction of decoys recovered) for the headline receptor sources of each target, plotted on the logarithmic decoy axis over which adjusted LogAUC is integrated (Fig.~\ref{si:logauc_explainer}); the dotted line marks the $1\%$ EF cutoff and the shaded region under the dashed diagonal is the zero-skill random baseline. \textbf{(a)}~JAK2 and \textbf{(b)}~ADRB2. In each panel the reference ligand-bound crystal used as the DUD-E template (3LPB for JAK2, 3NY8 for ADRB2) is drawn as a solid blue line and the remaining ligand-bound crystals as a shaded blue band spanning their per-point min--max range; for JAK2 the apo crystal (29RY) is shown as a separate dashed grey line. These are compared against the RRF-consensus rankings for AF2RAVE (green) and for AF2 top-pLDDT (orange), the same consensus receptors quantified in Fig.~\ref{fig:enrichment} and Fig.~\ref{fig:adrb2}E. For JAK2 the two sequence-based consensus rankings track within the crystal band across the whole curve, and for ADRB2 they sit at the upper edge of the band, including in the early-enrichment region left of the $1\%$ cutoff. These curves suggest that confidence-score-based curated AF2 and AF2RAVE ensembles recover actives comparably to---and for ADRB2 slightly ahead of---the experimental ligand-bound structures, without any ligand-bound template. Each curve is built with the pipeline's own score-tie-aware ROC construction and is therefore exactly consistent with the per-receptor AUC, adjusted LogAUC, and EF$_{1\%}$ values reported in the main text.}
    \label{si:roc}
\end{figure}

\begin{table*}[htbp]
    \centering
    \caption{\textbf{DOCK6 docking parameters, identical across every benchmark receptor.} The sphere-selection (sphgen), grid-generation, and flexible-docking settings below were the same for all receptors in both benchmarks---the JAK2 and ADRB2 ligand-bound crystals, the AF2RAVE-selected conformations, and the AF2 models alike---so that enrichment differences between receptors reflect their conformation and not the docking protocol.}
    \label{si:docking_params}
    \footnotesize
    \begin{tabular}{@{}llll@{}}
    \toprule
    Parameter & Value & Parameter & Value \\
    \midrule
\addlinespace
\multicolumn{4}{@{}l}{\textbf{Sphere selection (sphgen, \texttt{INSPH})}}\\
\texttt{sph\_dotlim} & 0.0 & \texttt{sph\_radmin} & 1.4 \\
\texttt{sph\_method} & R & \texttt{sph\_subset} & X \\
\texttt{sph\_radmax} & 4.0 &  &  \\
\addlinespace
\multicolumn{4}{@{}l}{\textbf{Grid generation (\texttt{grid.in})}}\\
\texttt{allow\_non\_integral\_charges} & yes & \texttt{dielectric\_factor} & 4 \\
\texttt{atom\_model} & a & \texttt{distance\_dielectric} & yes \\
\texttt{attractive\_exponent} & 6 & \texttt{energy\_cutoff\_distance} & 10 \\
\texttt{bump\_filter} & yes & \texttt{energy\_score} & yes \\
\texttt{bump\_overlap} & 0.75 & \texttt{grid\_spacing} & 0.3 \\
\texttt{compute\_grids} & yes & \texttt{output\_molecule} & no \\
\texttt{contact\_score} & no & \texttt{repulsive\_exponent} & 9 \\
\addlinespace
\multicolumn{4}{@{}l}{\textbf{Flexible ligand docking (\texttt{flex.in})}}\\
\texttt{atom\_model} & all & \texttt{pruning\_orient\_score\_cutoff} & 1000.0 \\
\texttt{automated\_matching} & yes & \texttt{rank\_ligands} & no \\
\texttt{bump\_filter} & no & \texttt{score\_molecules} & yes \\
\texttt{calculate\_rmsd} & no & \texttt{score\_threshold} & 100.0 \\
\texttt{chemical\_matching} & no & \texttt{simplex\_anchor\_max\_iterations} & 1000 \\
\texttt{conformer\_search\_type} & flex & \texttt{simplex\_cycle\_converge} & 1.0 \\
\texttt{contact\_score\_primary} & no & \texttt{simplex\_final\_min} & no \\
\texttt{critical\_points} & no & \texttt{simplex\_grow\_max\_iterations} & 1000 \\
\texttt{grid\_lig\_efficiency} & no & \texttt{simplex\_grow\_tors\_premin\_iterations} & 0 \\
\texttt{grid\_score\_es\_scale} & 1 & \texttt{simplex\_max\_cycles} & 200 \\
\texttt{grid\_score\_primary} & yes & \texttt{simplex\_random\_seed} & 0 \\
\texttt{grid\_score\_rep\_rad\_scale} & 1 & \texttt{simplex\_restraint\_min} & no \\
\texttt{grid\_score\_vdw\_scale} & 1 & \texttt{simplex\_rot\_step} & 0.0873 \\
\texttt{internal\_energy\_cutoff} & 100.0 & \texttt{simplex\_score\_converge} & 0.1 \\
\texttt{internal\_energy\_rep\_exp} & 12 & \texttt{simplex\_tors\_step} & 10.0 \\
\texttt{max\_orientations} & 5000 & \texttt{simplex\_trans\_step} & 0.2 \\
\texttt{minimize\_anchor} & yes & \texttt{use\_advanced\_simplex\_parameters} & no \\
\texttt{minimize\_flexible\_growth} & yes & \texttt{use\_clash\_overlap} & no \\
\texttt{minimize\_flexible\_growth\_ramp} & yes & \texttt{use\_database\_filter} & no \\
\texttt{minimize\_ligand} & yes & \texttt{use\_internal\_energy} & yes \\
\texttt{num\_final\_scored\_poses} & 1 & \texttt{use\_ligand\_spheres} & no \\
\texttt{num\_preclustered\_conformers} & 10 & \texttt{use\_rmsd\_reference\_mol} & no \\
\texttt{orient\_ligand} & yes & \texttt{write\_growth\_trees} & no \\
\texttt{pruning\_conformer\_score\_cutoff} & 100.0 & \texttt{write\_mol\_solvation} & no \\
\texttt{pruning\_conformer\_score\_scaling\_factor} & 1.0 & \texttt{write\_orientations} & no \\
    \bottomrule
    \end{tabular}
\end{table*}

\begin{table*}[htbp]
    \centering
    \caption{\textbf{AutoDock Vina docking parameters, identical across all JAK2 benchmark receptors.} Companion to the DOCK6 settings in Table~\ref{si:docking_params}, for the Vina enrichment of SI Fig.~\ref{si:vina_enrichment}. Every JAK2 Vina run used the same search-box dimensions and search/scoring settings (AutoDock Vina CPU); the box was centered on each receptor's ATP pocket, so its center differs per receptor and is omitted. Box dimensions are in \AA{}.}
    \label{si:vina_params}
    \setlength{\tabcolsep}{9pt}
    \renewcommand{\arraystretch}{1.3}
    \begin{tabular}{@{}ccc@{\hskip 3.5em}cccccc@{}}
    \toprule
    \multicolumn{3}{c}{\textbf{Search box (\AA)}} & \multicolumn{6}{c}{\textbf{Search and scoring}} \\
    \cmidrule(r{2.5em}){1-3} \cmidrule(l){4-9}
    \texttt{size\_x} & \texttt{size\_y} & \texttt{size\_z} & \texttt{exhaustiveness} & \texttt{num\_modes} & \texttt{min\_rmsd} & \texttt{energy\_range} & \texttt{scoring} & \texttt{cpu} \\
    \midrule
    25 & 25 & 25 & 40 & 10 & 3 & 3 & vina & 32 \\
    \bottomrule
    \end{tabular}
\end{table*}

\begin{figure*}[htbp]
    \centering
    \includegraphics[width=\textwidth]{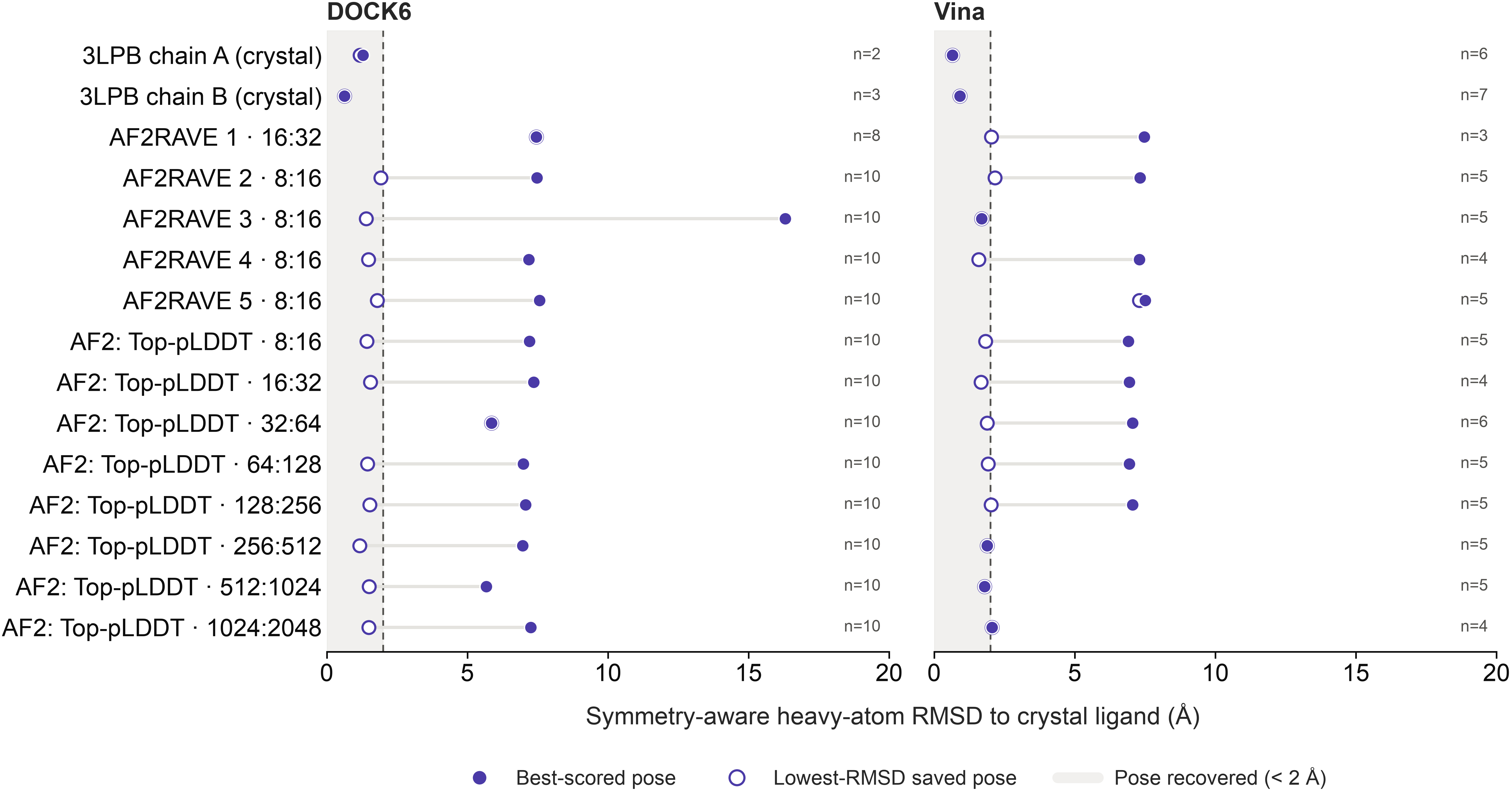}
    \caption{\textbf{Both docking engines reproduce the JAK2 crystal pose in the pipeline.} Self-redocking of the co-crystallized JAK2 ligand (NVB) as a functional check that the DOCK6 and AutoDock Vina (CPU) engines run end to end within ANTIOCH-VS; this is a pipeline sanity check, not an evaluation or comparison of the engines' redocking accuracy. For each receptor, the symmetry-aware heavy-atom RMSD of the best-scored pose (filled) and of the lowest-RMSD saved pose (open) to the crystal ligand are shown, joined by a stem; the shaded band marks the conventional $<2$~\AA{} pose-recovery region, and $n$ is the number of saved poses. Rows: the 3LPB crystal (chains A and B), the five AF2RAVE-selected models, and the eight top-pLDDT AF2 models (labelled by reduced-MSA depth). Redocking into the 3LPB crystal recovers the pose within $2$~\AA{} for both engines; the sequence-derived model receptors represent a different backbone conformation than the 3LPB crystal, so the crystal ligand is not expected to reproduce its 3LPB pose in them. Engine settings are the DOCK6 parameters of SI Table~\ref{si:docking_params} and AutoDock Vina (CPU; \texttt{exhaustiveness} 40).}
    \label{si:redocking}
\end{figure*}

\begin{figure*}[htbp]
    \centering
    \includegraphics[width=\textwidth]{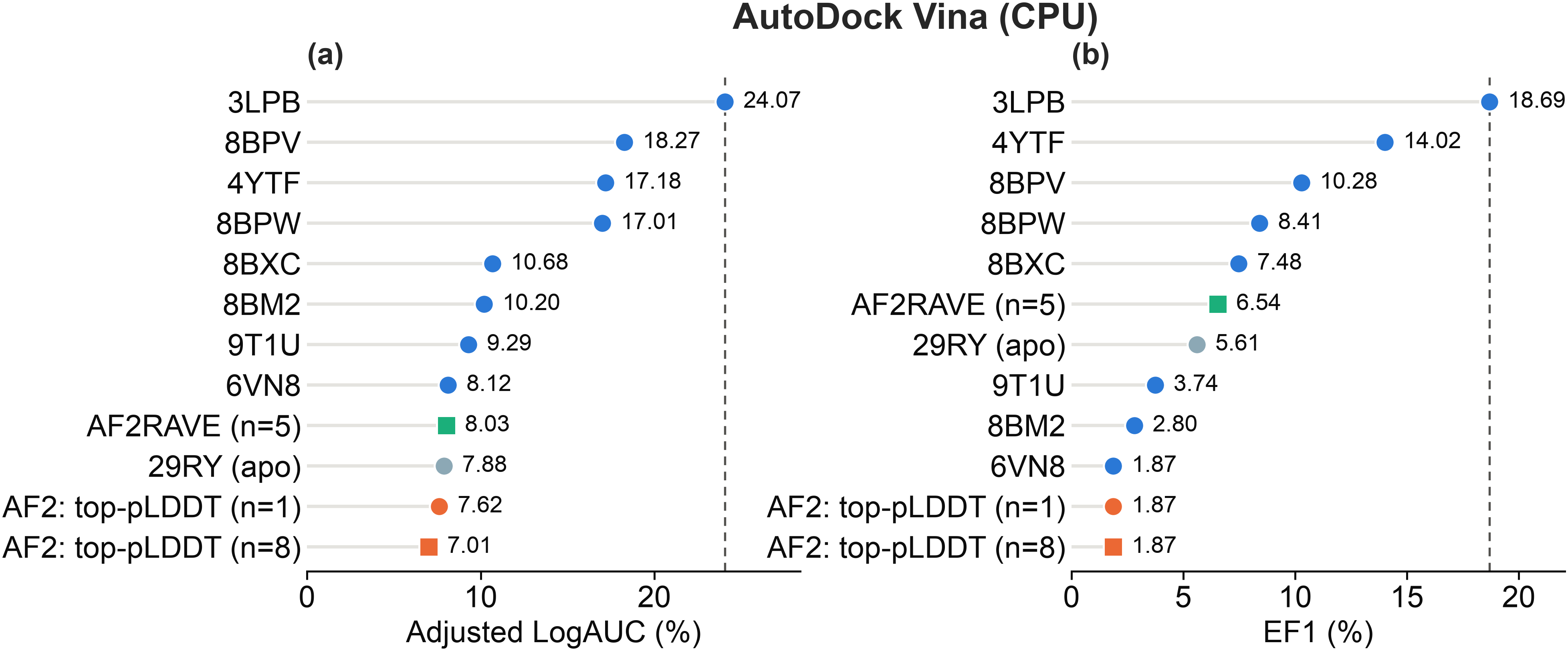}
    \caption{\textbf{JAK2 enrichment under AutoDock Vina.} The per-receptor JAK2 DUD-E enrichment of Fig.~\ref{fig:enrichment}, recomputed from AutoDock Vina (CPU) scores, showing that the pipeline runs end to end with Vina as well as DOCK6; DOCK6 is the main-text engine and this panel is provided for completeness, not as an evaluation or ranking of the two engines. \textbf{(a)}~Adjusted LogAUC and \textbf{(b)}~EF1 per receptor, with the 3LPB ligand-bound crystal as the dashed reference; crystals and the AF2RAVE consensus are ranked together by each metric, with the two structure-free AF2 rows pinned below. Colours and markers follow Fig.~\ref{fig:enrichment} (crystal, blue; apo, grey; AF2, orange; AF2RAVE, green; square, consensus/ensemble; circle, single structure). As under DOCK6, the AF2RAVE-selected ensemble enriches above the structure-free alternatives (the single AF2 model, the AF2 consensus, and the apo crystal) on both metrics. All Vina receptors were docked with identical settings (box $25^3$, \texttt{exhaustiveness} 40, \texttt{scoring} vina).}
    \label{si:vina_enrichment}
\end{figure*}

\begin{figure}[htbp]
    \centering
    \includegraphics[width=\columnwidth]{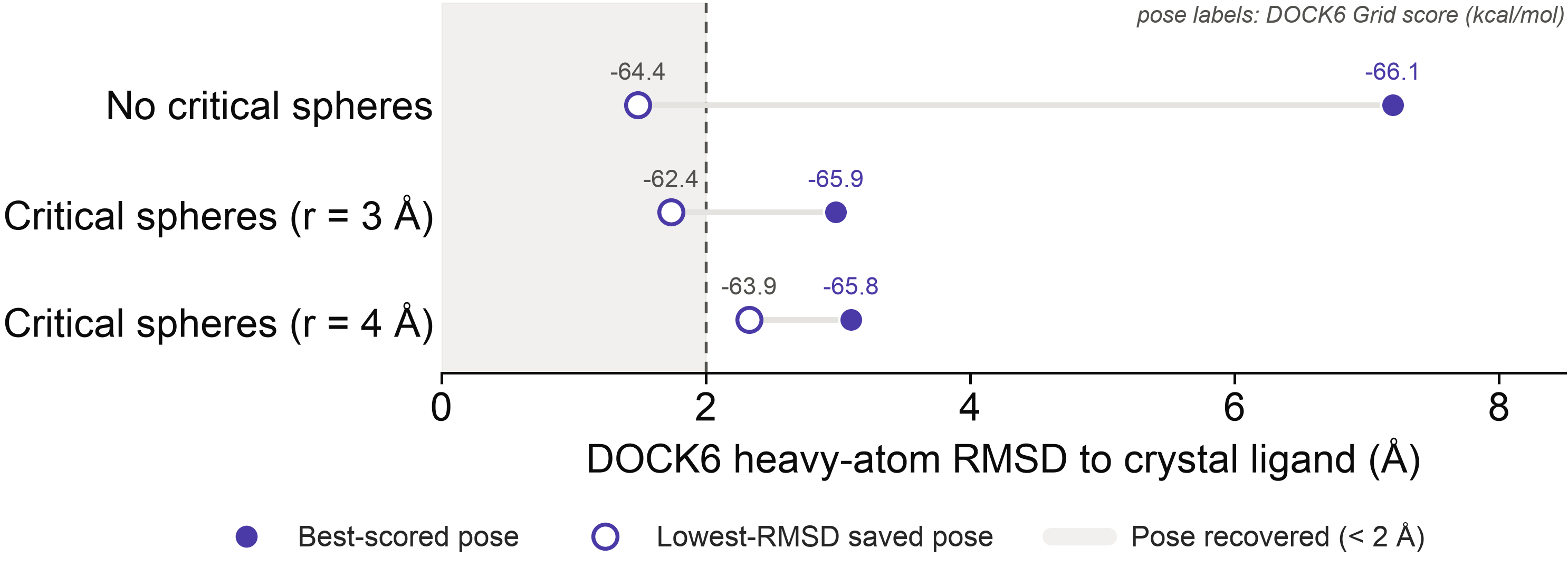}
    \caption{\textbf{DOCK6 critical spheres: configurable site restriction (illustrative).} ANTIOCH-VS can designate ``critical'' spheres that a docked pose must occupy, restricting docking to a user-specified sub-pocket. Here, for one AF2RAVE-selected JAK2 structure, critical spheres were placed at the kinase hinge (GLU93/LEU95 backbone) and the crystal ligand was redocked with the restriction off (no critical spheres), on at sphere radius $3$~\AA{}, and on at $4$~\AA{}. Per condition, the best-scored pose (filled) and the lowest-RMSD saved pose (open) are shown by DOCK6's reported heavy-atom RMSD (HA\_RMSD) to the crystal ligand; the shaded band is the conventional $<2$~\AA{} recovery region, and each pose is labelled with its DOCK6 Grid score (kcal/mol). Without the restriction, DOCK6 samples a near-native hinge pose ($\sim$1.5~\AA{}) but top-scores an off-hinge pose ($\sim$7~\AA{}); enabling the hinge critical spheres makes the near-native pose the best-scored one ($\sim$3~\AA{}) at negligible Grid-score cost. This is a single illustrative example of the capability, not a systematic evaluation; the effect of critical spheres on enrichment was not assessed in this work.}
    \label{si:critical_spheres}
\end{figure}

\end{appendices}

\clearpage  
\bibliography{antioch_references}

\end{document}